\documentclass[aps,pra,twocolumn,showpacs,superscriptaddress%,longbibliography
]{revtex4-1}
\usepackage{graphicx}
\usepackage{amsmath}
\usepackage{amssymb}
\usepackage{epstopdf}
\usepackage{amsfonts}
\usepackage{braket}
\usepackage{dcolumn}
\DeclareMathOperator{\Tr}{Tr}
\usepackage{dcolumn}
\newcolumntype{d}[1]{D{.}{.}{#1}}
\begin{document}
\title{
Temperature-dependence of
small harmonically trapped atom systems
with Bose, Fermi and Boltzmann statistics
}
\author{Yangqian Yan}
\affiliation{Department of Physics and Astronomy,
Washington State University,
  Pullman, Washington 99164-2814, USA}
\author{D. Blume}
\affiliation{Department of Physics and Astronomy,
Washington State University,
  Pullman, Washington 99164-2814, USA}
\date{\today}

\begin{abstract}
While the zero-temperature properties of harmonically
trapped cold few-atom systems have been discussed fairly 
extensively over the past decade, much less is known about
the finite-temperature properties. Working in the canonical ensemble,
we characterize small harmonically trapped
atomic systems as a function of the
temperature using analytical and numerical techniques.
We present results for the energetics, structural properties,
condensate fraction, superfluid fraction, and superfluid density.
Our calculations for the two-body system underline that the
condensate and superfluid fractions are distinctly different
quantities. 
Our work demonstrates that the path integral Monte Carlo method
yields reliable results for bosonic and fermionic systems over a
wide temperature range, including the regime where the de Broglie 
wave length is large, i.e., where the statistics plays an
important role. The regime where the Fermi sign problem leads to
reasonably large signal to noise ratios is mapped out for selected
parameter combinations. Our calculations for bosons focus on the 
unitary regime, where the physics is expected to be governed by
the three-body parameter. If the three-body parameter is large compared to the
inverse of the harmonic oscillator length, we find that the bosons form a droplet
at low temperature and behave approximately like a non-interacting
Bose and eventually Boltzmann gas at high
temperature. The change of the behavior occurs over a fairly 
narrow temperature range. A simple model that reproduces the
key aspects of the phase transition like feature, which can potentially
be observed in cold atom Bose gas experiments, is presented.
\end{abstract}
\maketitle

\section{Introduction}
\label{sec_intro}
Ultracold atomic gases 
provide a flexible
platform for studying a myriad of phenomena
that are driven by quantum 
mechanics~\cite{stringariBosereview,pethickbook,stringaribook,blochreview,stringariFermireview,blumerev12}.
Generally speaking, quantum statistical effects dominate when
the de Broglie wave length is comparable to or larger than the
average interparticle spacing.
When the de Broglie wave length
is small, 
the particle statistics 
plays a negligible role and the system dynamics is
governed by Boltzmann statistics.
Since the de Broglie wave length scales as 
$1/\sqrt{T}$~\cite{stringaribook,pethickbook}, where
$T$ is the temperature, changing the temperature allows
one to turn the particle statistics
``on'' and ``off''.
Atomic gases, which can be cooled to below the
quantum degeneracy temperature, thus provide
an ideal platform for investigating the importance
of particle statistics.

For macroscopic samples, a prominent example for  
a thermal phase transition is
the transition from the normal to the 
superfluid phase as observed in
bosonic liquid $^4$He and fermionic liquid 
$^3$He~\cite{book_on_superfluids_and_superconductors}. 
Bose-Einstein condensation, the macroscopic occupation of
a single particle state, is another important example.
While Bose-Einstein condensation occurs
for ultracold  bosonic atomic gases~\cite{stringariBosereview}, it does not
occur, at least not directly,
for ultracold fermionic atomic gases~\cite{stringariFermireview,leggettbook}.
Condensation for fermions occurs only when 
two fermions form 
composite bosons (diatomic molecules or Cooper 
pairs)~\cite{stringariFermireview,leggettbook,experimentalPaperByJin,experimentalPaperByKetterleGroup}.
If the number of particles is
finite (as opposed to infinite), phase transitions get smeared out
and the usual concept, which considers statistical properties
in the thermodynamic limit,
has to be revised~\cite{schunckbook,pathria1996statistical}.

The main objective of this paper is to study the temperature dependence
of small harmonically trapped atomic Bose and Fermi systems.
To describe these systems, we adopt the canonical ensemble, 
i.e., we assume that the system under study is in 
thermal contact with a heat bath or thermostat,
which has a large number of particles
and a well defined temperature $T$~\cite{thermobasicText}.
We monitor various system properties as a function of the
temperature, the number of particles,
the particle statistics, and the interaction
strength. 
Particular emphasis is placed on the strongly-interacting
unitary regime, where the $s$-wave scattering length
diverges. At zero temperature, it is well established
that the particle statistics has a paramount effect on the
system properties. 
Two-component Fermi gases with infinitely
large interspecies scattering length are fully described
by the $s$-wave scattering length 
alone~\cite{stringariFermireview,blochreview,petrov,earlyDMCworkByCarlson,earlyDMCworkByGiorgini}, while the properties
of Bose gases additionally depend on a three-body 
parameter~\cite{Efimov,BraatenHammerReview}. 
These fundamental differences, which are due to the particle
statistics, continue to play an important role at low
temperature but die out at sufficiently high temperature.
An interesting question, which we attempt
to answer in this paper, is thus what happens at intermediate
temperatures.
As expected, we find that the low and intermediate temperature
behavior of Bose and Fermi gases is vastly different.
For certain parameter combinations, 
we find a thermal phase transition like feature for Bose systems that is
governed by the three-body Efimov parameter.
Specifically, we find a transition from a droplet like state
to a gas-like state.
No such transition exists for two-component Fermi gases.

The remainder of this paper is organized as follows.
Section~\ref{sec_theory} introduces the system Hamiltonian
and reviews the connections between the
free-space Efimov spectrum and the zero
temperature spectrum of the harmonically
trapped three-boson system. Moreover, the path integral Monte Carlo (PIMC) approach
is introduced and some numerical details are discussed.
Section~\ref{sec_twobody} presents finite temperature
characteristics of the trapped two-atom system. Emphasis
is placed on the condensate and superfluid fractions.
The radial superfluid density is calculated and analyzed.
Section~\ref{sec_nbody} discusses our 
finite temperature results for systems with three and 
more particles.
Section~\ref{sec_boson} 
focuses on systems consisting of $N$ identical bosons while
Sec.~\ref{sec_mixture}
considers a trapped gas with Bose, Fermi or Boltzmann statistics 
with an impurity.
Lastly, Sec.~\ref{sec_conclusion} concludes.

\section{Theoretical Background}
\label{sec_theory}
\subsection{System Hamiltonian and observables}
\label{SecObservables}
This section introduces the system Hamiltonian and reviews two frameworks
for determining thermally averaged observables.
We fix the number of particles and work in the
canonical ensemble.
We consider $N$ particles with position vectors $\mathbf{r}_j$
and mass $m_{\text{a}}$
in a
spherically symmetric harmonic trap with angular trapping frequency
$\omega$.
The model Hamiltonian $\hat{H}$ reads
\begin{equation}\label{GeneralHamiltonian}
  \hat{H}= \hat{H}_0+ \hat{V},
\end{equation} 
where $\hat{H}_0$,
\begin{equation}\label{NonInteractingHamiltonian}
  \hat{H}_0=
  \sum_{j=1}^{N}\left( \frac{-\hbar^2}{2m_{\text{a}}}\nabla_j^2+
  \frac{1}{2}m_{\text{a}}
  \omega^2\mathbf{r}_{j}^2
  \right),
\end{equation} 
denotes the non-interacting Hamiltonian.
The interaction potential $\hat{V}$ reads
\begin{equation}
  \hat{V}=\sum_{j=1}^N\sum_{k>j}^N V_{\text{tb}}^{(jk)}(r_{jk}),
  \label{Interaction}
\end{equation}
where ${r}_{jk}$ ($r_{jk}=|\mathbf{r}_j-\mathbf{r}_k|=|\mathbf{r}_{jk}|$)
denotes the relative distance between the $j$th and 
$k$th particles and
$V_{\text{tb}}^{(jk)}$ 
the interaction potential for 
the $j$th and $k$th particles.
We employ two different interaction models.
Our calculations presented in Sec.~\ref{sec_twobody} employ
the regularized zero-range Fermi-Huang 
pseudopotential $V_{\text{F}}^{(jk)}$~\cite{Yang1957}
with $s$-wave scattering length $a_s^{(jk)}$.
Our PIMC calculations presented in Sec.~\ref{sec_nbody}
employ 
a finite-range Gaussian potential 
$V_{\text{G}}^{(jk)}$, where
$V_{\text{G}}^{(jk)}({r}_{jk})=
U_{0}^{(jk)} \exp 
[ -
{r}_{jk}^2/(2r_{0}^2)
]$ with depth $U_0^{(jk)}$ ($U_0^{(jk)}<0$) and
range $r_0$.
The depth and range are adjusted so that $V_{\text{G}}^{(jk)}$
yields the desired $s$-wave scattering length $a_s^{(jk)}$.
Throughout, we consider potentials
that support at most one
free-space 
$s$-wave bound state and whose range $r_0$ is much smaller than the
characteristic harmonic trap length $a_{\text{ho}}$,
where $a_{\text{ho}}=\sqrt{\hbar/(m_{\text{a}} \omega)}$.

To calculate thermally averaged quantities, we introduce the density
operator $\hat{\rho}$~\cite{pathria1996statistical,CeperleyRMP},
\begin{equation}
  \hat{\rho}=e^{-\beta \hat{H}},
  \label{densityoperatororigin}
\end{equation}
where $\beta$ is the inverse temperature, $\beta=1/(k_B T)$.
The expectation value for an operator $\hat{\mathcal{O}}$ is
$\Tr(\hat{\rho}\hat{\mathcal{O}})/Z$, where ``$\Tr$'' stands for the trace of the
matrix that is created by projecting the operator onto a
complete basis set, and $Z=\Tr(\hat{\rho})$ is the partition function.

A convenient basis set consists of the energy 
eigen states $\psi_j$ of the Hamiltonian $\hat{H}$. In this case, the
density operator is diagonal and can be 
written as~\cite{pathria1996statistical}
\begin{equation}
  \hat{\rho}=\sum_j e^{-\beta E_j} \Ket{\psi_j}\Bra{\psi_j},
  \label{densityoperator}
\end{equation}
where $E_j$ denotes the eigen energy of state $\psi_j$,
and the partition function reads 
\begin{eqnarray}
\label{eq_Znonsymm}
Z=\sum_j e^{-\beta E_j}.
\end{eqnarray}
The sums in Eqs.~(\ref{densityoperator}) and (\ref{eq_Znonsymm})
are limited to the energy eigen states  $\psi_j$ that have the
proper particle statistics.
For $N=2$, e.g., the eigen states can be grouped into states that are symmetric and those that are anti-symmetric under the exchange of the
two particles.
If we treat two identical bosons (fermions),
only the symmetric (anti-symmetric) states are included 
in the sums in Eqs.~(\ref{densityoperator}) and (\ref{eq_Znonsymm}).
Importantly, if the complete set is known, the thermal average 
$\langle \hat{\mathcal{O}} \rangle$ of the operator $\hat{\mathcal{O}}$
can be calculated,
\begin{equation}\label{ThermalAverageEigenFunction}
  \langle \hat{\mathcal{O}}
  \rangle=Z^{-1}\sum_je^{-\beta E_j}\langle \psi_j|\hat{\mathcal{O}}|\psi_j
\rangle.
\end{equation} 
While the determination of a large number of energy eigen states
$\psi_j$ is feasible for small systems, say $N \lesssim 4$,
it becomes unfeasible for larger systems.

An alternative formulation, 
which forms the starting point of the PIMC
approach~\cite{CeperleyRMP} (see Sec.~\ref{secPIMC} for details),
projects the density operator onto the position basis,
$\rho_{\text{non-symm}}(\mathbf{R},\mathbf{R}',\beta)=\Braket{\mathbf{R}|\hat{\rho}|\mathbf{R}'}$.
Here, $\mathbf{R}$ and $\mathbf{R}'$
collectively denote the position vectors
$\mathbf{r}_1,\cdots,\mathbf{r}_N$ and 
$\mathbf{r}'_1,\cdots,\mathbf{r}'_N$, respectively.
The thermal average of the operator $\hat{\mathcal{O}}$
then reads
\begin{eqnarray}\label{ThermalAverage}
  \langle \hat{\mathcal{O}}
  \rangle_{\text{non-symm}}=
(Z_{\text{non-symm}})^{-1} \times \nonumber \\
\int\mathrm{d}\mathbf{R}\mathrm{d}\mathbf{R}'\rho_{\text{non-symm}}(\mathbf{R},\mathbf{R}',\beta)\langle\mathbf{R}'|\hat{\mathcal{O}}|\mathbf{R} \rangle,
\end{eqnarray} 
where the partition function $Z_{\text{non-symm}}$,
\begin{equation}\label{}
  Z_{\text{non-symm}}=
\int \mathrm{d}\mathbf{R}\rho_{\text{non-symm}}(\mathbf{R},\mathbf{R},\beta),
\end{equation}
is again the trace over the diagonal elements.
To properly symmetrize or anti-symmetrize the density
operator, we introduce 
the symmetrizer $\hat{\mathcal{P}}$~\cite{CeperleyRMP}. 
For the single-component Bose and Fermi
gases, $\hat{\mathcal{P}}$ can
be written as~\cite{basicTextBaym}
\begin{equation}
\hat{\mathcal{P}}=
\frac{1}{N!}\sum_{\sigma}(\pm1)^{N_\text{I}(\sigma)} \hat{P}_{\sigma},
  \label{}
\end{equation}
where $\sigma$ denotes
the permutation of particle indices, $N_\text{I}(\sigma)$ the
number of inversions in $\sigma$~\footnote{In {\it{Mathematica}},
the number of inversions $N_\text{I}(\sigma)$ is represented by the function Inversions}, 
and $\hat{P}_{\sigma}$ the corresponding
permutation operator.
For two identical fermions, e.g., $\hat{{\mathcal{P}}}$ reads
$(1-\hat{P}_{12})/2$, where $\hat{P}_{12}$ exchanges 
the particle labels $1$ and $2$.
For mixtures, the symmetrizer $\hat{\mathcal{P}}$
has to be generalized appropriately.
The partition function and thermally averaged observables then
read~\cite{CeperleyRMP}
\begin{eqnarray}
  Z=
\int \mathrm{d}\mathbf{R}\rho(\hat{\mathcal{P}}\mathbf{R},\mathbf{R},\beta)
\end{eqnarray}
and
\begin{eqnarray}
\label{eq_ThermalAverageSymm}
  \langle \hat{\mathcal{O}}
  \rangle=Z^{-1}\int\mathrm{d}\mathbf{R}\mathrm{d}\mathbf{R}'\rho(\hat{\mathcal{P}}\mathbf{R},\mathbf{R}',\beta)\langle \mathbf{R}'|\hat{\mathcal{O}}|\mathbf{R} \rangle.
  \label{<++>}
\end{eqnarray}

In addition to the thermally averaged energy $E$, this work considers a
number of thermally averaged structural properties.
The scaled radial density $4 \pi r_j^2 \rho_{\text{rad}}(r_j)$ 
with normalization $4\pi \int dr_j \rho_{\text{rad}}(r_j)
r_j^2=N$~\footnote{For the two-component system, normalization to the number
of particles in each component would be a more natural choice.}
tells one the likelihood of finding the $j$th particle at distance $r_j$ from
the trap center.
The scaled
pair distribution function $4 \pi r_{jk}^2 P_{\text{pair}}(r_{jk})$ with
normalization $4\pi\int dr_{jk} P_{\text{pair}}(r_{jk}) r_{jk}^2=1$
tells one the likelihood of finding particles $j$ and $k$ at distance
$r_{jk}$.
The 
hyperradial distribution function 
$P_{\text{hyper}}(R)$ with normalization 
$\int dR P_{\text{hyper}}(R) =1$
tells one the likelihood of finding particles $j$, $k$ and $l$
in a configuration of size $R$; here, $R^2=(r_{jk}^2+r_{kl}^2+r_{jl}^2)/3$.
For $N=3$, $R$ is the hyperradius (see Sec.~\ref{SecEfimov} for details).

We also consider the
condensate fraction, superfluid fraction, and superfluid density.
For homogeneous systems, the
condensate fraction $n_c$ is typically defined
through the large distance behavior of 
the one-body density matrix for bosons and 
the two-body density matrix for two-component 
fermions~\cite{stringariFermireview,leggettbook}.
It indicates the off-diagonal long-range order of the system.
For inhomogeneous systems, the condensate fraction is
defined as the largest eigen value of the
one- and two-body density matrices for bosons and fermions,
respectively~\cite{famousWorkByOnsager,leggettbook,LookAtTheReferencesCitedByCederbaumsGroup}.
Intuitively, it is clear that the long-range behavior is 
``cut off'' by the confinement or the finite extend of the
system, implying that the asymptotic behavior of the density matrix
contains no information about non-trivial correlations.
Section~\ref{sec_twobody} reports the dependence of 
the condensate fraction $n_c$ on the temperature for two identical bosons and two
distinguishable particles.
These studies extend
the zero temperature calculations of $n_c$ presented in
Ref.~\cite{KevinDesmondPRA} to finite temperature.
The finite temperature behavior of $n_c$
has previously been reported for two harmonically
trapped particles in one dimension~\cite{JPBpaperIncludesWilkensAsAnAuthor}
but not, to the best of our knowledge, for two harmonically trapped particles 
in three dimensions.

The superfluid fraction 
$n_s$ can be defined 
in various ways (see, e.g., Refs.~\cite{PaperThatIPointedOutByBrand,ViewPointByCarusotto,leggettbook,farrelly14} for a discussion).
In this work, we utilize the 
moment of inertia based definition,
which has its origin in the
two-fluid model~\cite{leggett70,baym69,leggett73,ceperley87},
\begin{eqnarray}
\label{eq_superfluidfraction}
n_s=1- \frac{I_{\text{q}}}{I_{\text{c}}}.
\end{eqnarray}
The quantum moment of inertia $I_{\text{q}}$ is 
defined in terms of the response
to an infinitesimal rotation about the $z$-axis,
\begin{eqnarray}
I_{\text{q}}=\beta (\langle \hat{L}_{\text{tot},z}^2\rangle- 
\langle \hat{L}_{\text{tot},z} \rangle ^2),
\label{eq_quantummomentofinertia}
\end{eqnarray}
where $\hat{L}_{\text{tot},z}$ denotes the $z$-component of the
total angular momentum operator $\hat{\mathbf{L}}_{\text{tot}}$.
The classical moment of inertia $I_{\text{c}}$ is defined 
through 
\begin{eqnarray}
  I_{\text{c}}=\langle m_{\text{a}}\sum_{j} r_{j,\perp}^{2} \rangle,
\end{eqnarray}
where  $r_{j,\perp}$ denotes the distance
of the
$j$th particle to the $z$-axis,
$r_{j,\perp}=| \mathbf{r}_j \times \hat{z}|$.
The superfluid density is defined such that
$m_{\text{a}} \int d \mathbf{r}  \rho_s(\mathbf{r}) r_{\perp}^2=I_{\text{c}}-I_{\text{q}}$,
where $r_{\perp}$ denotes the distance to the $z$-axis~\cite{kwon06}. 
The moment of inertia based definitions of the superfluid
fraction and superfluid density have previously been applied to a variety of finite-sized quantum liquids~\cite{grebenev98,grebenev00,tang02,kwon02,kwon05,boninsegnipra07,boninsegniprl08}.
Knowing the complete set of energy eigen states
and eigen energies and using Eq.~(\ref{ThermalAverageEigenFunction}), the 
thermally averaged expectation values 
$\langle \hat{L}_{\text{tot},z} \rangle$ 
and
$\langle \hat{L}_{\text{tot},z}^2 \rangle$ 
can be calculated, thereby yielding $I_{\text{q}}$.
Within the PIMC approach, the superfluid fraction and superfluid
radial density are evaluated using the area 
estimator~\cite{ceperley87,ceperley89,ceperley03,kwon06}
(see Sec.~\ref{secPIMC} for details on the PIMC approach).

\subsection{Efimovian states of three identical bosons in a trap}
\label{SecEfimov}
This section reviews the zero-temperature properties of 
three identical harmonically-trapped bosons.
As discussed in the literature~\cite{blumerev12}, 
harmonically trapped unitary Bose and Fermi
gases with short-range interactions 
exhibit universal properties,
provided the range 
of the interaction is smaller than all other length scales in the
problem.
The properties of the two-component Fermi gas near a broad
$s$-wave resonance (and away
from $p$- and higher-partial wave resonances) are governed by 
the interspecies $s$-wave scattering length $a_s$ and the harmonic oscillator
length $a_{\text{ho}}$.
In the unitarity limit, i.e., 
for $|a_s|=\infty$, the $s$-wave scattering length
does not define a 
meaningful length scale and the only remaining length scale is
$a_{\text{ho}}$~\cite{stringariFermireview,blochreview,blumerev12,wernercastinprl,wernercastinpra}.
The corresponding energy scale is $E_{\text{ho}}=\hbar \omega$.
For three or more identical bosons, an additional parameter, namely,
the three-body parameter
$\kappa_*$, is needed to describe the ground state properties of the
Bose
gas~\cite{Efimov,BraatenHammerReview,wernercastinprl,wernercastinpra}.

The role of $\kappa_*$ 
can be made transparent using  the hyperspherical coordinate
approach~\cite{blumerev12,BraatenHammerReview}.
To this end,
we separate off the center of mass motion and divide
the remaining six coordinates into the hyperradius $R$ 
and five hyperangles collectively denoted by $\mathbf{\Omega}$.
In the limit of pairwise 
additive zero-range interactions with $1/a_s=0$, the hyperradial and
hyperangular degrees of freedom are 
separable~\cite{Efimov,BraatenHammerReview,wernercastinprl}.
The lowest eigen value of the hyperangular
Schr\"odinger equation for the channel with vanishing relative
angular momentum angular $l$ is typically denoted by
$s_0$, where
$s_0\approx 1.006 \imath  $~\cite{Efimov,BraatenHammerReview}.
The hyperradial Hamiltonian $\hat{H}_R$ can then be written as
\begin{equation}
 \hat{H}_R=\frac{-\hbar^2}{2 m_{\text{a}}} \frac{\partial^2}{\partial
 R^2}+\frac{1}{2} m_{\text{a}} \omega^2 R^2+
 \frac{\hbar^2(s_0^2-1/4)}{2 m_{\text{a}} R^2}.
 \label{<++>}
\end{equation}
The last term can be interpreted
as an effective attractive potential, which diverges in the
$R=0$ limit.
Without a three-body parameter,
the system exhibits the Thomas collapse~\cite{thomas35}.
The scaled radial solution 
in the small $R$ limit 
is proportional to
$\sqrt{R}\sin(\operatorname{Im}(s_0) \ln R+\theta_b)$~\cite{JonsellPRL}, 
where 
$\theta_b$,
\begin{equation}
\label{eq_threebodyphase}
\theta_b=\arg\left( \frac{\Gamma(\frac{1}{2}-\frac{E_{\text{rel}}}{2E_{\text{ho}}}+\frac{s_0}{2})}{\Gamma(1+s_0)} \right),
\end{equation}
is the three-body phase that determines the short-range
behavior of the hyperradial wave function and $E_{\text{rel}}$ denotes the
relative three-body energy.
The three-body phase can be related to the three-body
parameter $\kappa_*$.

\begin{figure}
\centering
\includegraphics[angle=0,width=0.45\textwidth]{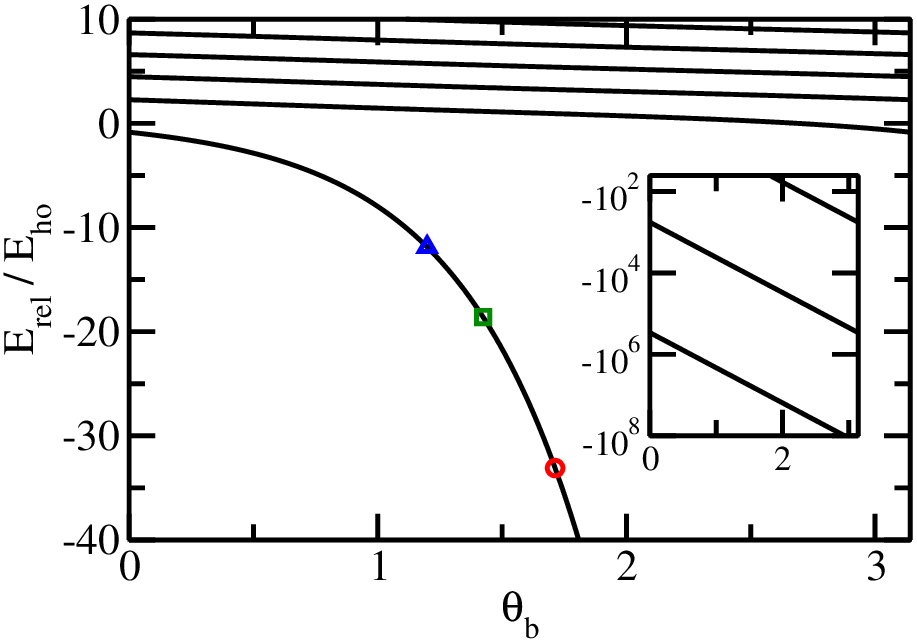}
\caption{(Color online)
Relative
energy spectrum as a function of the three-body phase $\theta_b$ 
  for three identical bosons in a harmonic trap interacting through 
  zero-range potentials with infinite $s$-wave scattering lengths. 
  The circle, square, and triangle show the ground state energy
  for the Gaussian two-body interaction with range
  $r_0/a_{\text{ho}}=0.06, 0.08,$ and $0.1$, respectively.
  The inset shows the negative energy regime on a log scale.
  The spacing between the energy levels for fixed $\theta_b$ is very close
to 515, i.e., very close to the free-space scaling factor.
 }\label{figthetab}
\end{figure}

Solving Eq.~(\ref{eq_threebodyphase}),
the solid lines in Fig.~\ref{figthetab} 
show the relative three-body zero-range eigen energies as a function of
the three-body phase $\theta_b$ 
for infinitely large $s$-wave scattering length. 
For a fixed $\theta_b$, the energies of the negative part of the 
energy spectrum are spaced roughly by the factor 
515~\cite{JonsellPRL,KokkelmansFBS}.
These geometrically spaced energy levels are the
signature of the three-body Efimov effect. 
In free space, the spacing is exactly 
$\exp(2 \pi /|s_0|) \approx (22.7)^2 \approx 515$ and
the three-body parameter $\kappa_*$ is defined as the binding momentum
of one of the Efimov trimers,
$E_{\text{rel}}=\hbar^2 \kappa_*^2/m$~\cite{BraatenHammerReview}. 
Knowing $\kappa_*$, the ratio between
consecutive energy levels of the free-space system is fixed.
For the trapped system, corrections arise when the trimer size
approaches the harmonic oscillator length.
For the states with positive energy, 
the spacing between consecutive
states is approximately 
$2E_{\text{ho}}$~\cite{JonsellPRL,KokkelmansFBS}~\footnote{Note that the range
of the three-body phase $\protect\theta_b$ in Fig.~9 of Ref.~\protect\cite{blumerev12} is incorrect:
The unique $\protect\theta_b$ values span a range of $\protect\pi$ instead of $2\protect\pi$}.

We now connect the energy spectrum 
for the Gaussian interaction model $V_{\text{G}}$
with that for the zero-range model.
In free-space, the three-body system with pairwise
Gaussian interaction supports 
infinitely many states. The spacing between the ground state and the first excited state 
at unitarity is $(22.98)^2$ and between
the energies of
the first excited state and the second excited state is $(22.7)^2$.
These values are close to the universal scaling factor. Indeed, the
Gaussian interaction model has been used extensively in the literature
to describe Efimov physics~\cite{Naidon,Kievsky,vonStecherJPB}.
For the trapped system, the ratio between the range $r_0$ of the two-body
interaction and the harmonic oscillator length comes into play.
The circle, square and triangle in Fig.~\ref{figthetab} show the 
relative energy of the lowest state of the trapped system
for $r_0/a_{\text{ho}}=0.06$, $0.08$ and $0.1$,
respectively.
Assuming that the zero-range energy spectrum provides a 
reasonable description,
Fig.~\ref{figthetab} allows us to estimate the three-body phase.

\begin{figure}
\centering
\includegraphics[angle=0,width=0.4\textwidth]{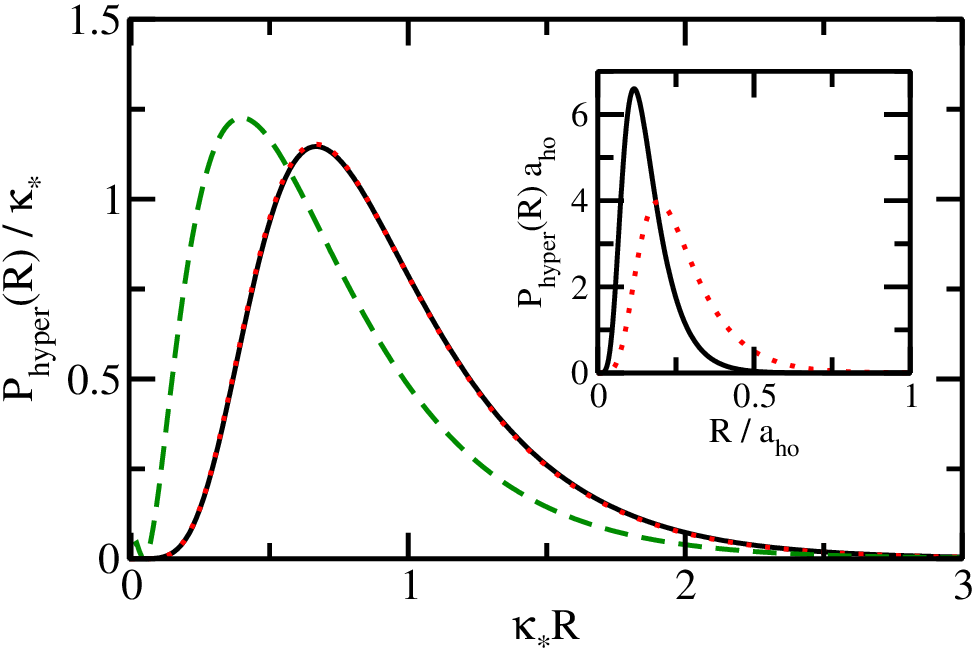}
\caption{(Color online)
  Hyperradial density $P_{\text{hyper}}(R)$ 
for three identical bosons at unitarity.
Solid and dotted lines show
  the PIMC results  at $k_BT/E_{\text{ho}}=0.4$
for the Gaussian model potential with
$r_0/a_{\text{ho}}=0.06$ and $0.1$, respectively
  (in the main panel, the curves are
  indistinguishable on the scale shown).
  The main panel and the inset show the same data but
  use a different scaling: The main panel
uses units derived from the energy
  of the three-boson system at $T=0$ 
 while the inset 
  employs harmonic oscillator units.
  For comparison, the dashed line shows the hyperradial density obtained 
using the
  zero-range pseudopotential
with $\kappa_*$ determined by the relative energy of the finite-range
potential.
 }\label{hyperradial}
\end{figure}

For our purposes, the size of the trimer compared to the range of the
interaction is relevant.
For the three $r_0$ considered, the size of the lowest trimer, as measured by
the expectation value of the hyperradius $R$, is roughly
$0.160a_{\text{ho}}$, $0.212 a_{\text{ho}}$ and 
$0.266 a_{\text{ho}}$, i.e., the trimers are much smaller than
$a_{\text{ho}}$, and thus very close to the free space trimers.
The lowest Efimov trimer is only a bit 
larger than $r_0$ (the size is about $2.66r_0$ for all cases), implying that we 
expect finite-range effects to be non-negligible.
Indeed, Fig.~\ref{hyperradial} shows that the hyperradial densities of the
lowest state of the 
finite-range three-body system (solid and dotted lines) differ notably 
from the hyperradial density of the zero-range system (dashed line).
This difference cannot be attributed to the fact that the hyperradial
densities are calculated at finite temperature (the finite-range $T=0$
hyperradial densities are,
on the scale chosen, indistinguishable from those
shown in Fig.~\ref{hyperradial}) but is due to finite-range effects.
Despite these finite-range corrections, 
the Gaussian interaction model
allows us to gain insights into finite-temperature effects that are
governed by the lowest Efimov state of the three-body system
(see Sec.~\ref{sec_nbody} for details).

\subsection{PIMC approach}
\label{secPIMC}
This section reviews the finite-temperature continuous-space PIMC
approach~\cite{CeperleyRMP}. 
The key idea behind the PIMC approach is to convert the calculations at 
low temperature (large $\beta$) into a series of calculations at
high temperature. 
Specifically, 
the PIMC approach
rewrites $\exp(-\beta \hat{H})$ in terms of
the product $\prod_{j=1}^M \exp(-\tau \hat{H})$, where
$\tau= \beta/M$.
The idea is to use a sufficiently small
$\tau$ (sufficiently large integer
$M$) so that the integrals involving $\tau$ can be factorized
with controllable error. In the calculations reported
in Sec.~\ref{sec_nbody}, we use $M \approx 400-7000$
(the actual number used depends on the
temperature $T$ and the two-body range $r_0$).
Inserting $\int_j\Ket{\mathbf{R}_j}\Bra{\mathbf{R}_j}$ repeatedly, 
Eq.~(\ref{eq_ThermalAverageSymm})
becomes~\cite{CeperleyRMP}
\begin{eqnarray}
  \langle \mathcal{\hat{O}}
  \rangle=Z^{-1}\int\mathrm{d}\mathbf{R}_0\dots\mathrm{d}\mathbf{R}_M\rho(\hat{\mathcal{P}}\mathbf{R}_0,\mathbf{R}_1,\tau)\times  \nonumber \\
\rho(\mathbf{R}_{1},\mathbf{R}_2,\tau)\times\dots \nonumber\\
  \times\rho(\mathbf{R}_{M-1},\mathbf{R}_M,\tau)\Braket{\mathbf{R}_M|\mathcal{O}|\mathbf{R}_0}.
  \label{observable}
\end{eqnarray} 
To evaluate expectation values of 
operators that probe the diagonal but not the off-diagonal 
elements of the 
real-space density matrix,
only closed paths with
$\hat{\mathcal{P}} \mathbf{R}_0 = \mathbf{R}_M$ are needed.
The density matrix
$\rho(\mathbf{R}_{j-1},\mathbf{R}_j,\tau)$
is, in general, unknown. 
Using the second- or fourth-order factorization~\cite{CeperleyRMP,chin97,voth}, 
the high-temperature 
density operator can be divided into the non-interacting and interacting
parts,
\begin{eqnarray}
  \exp \left[-\tau(\hat{H}_0+\hat{V}) \right]=
\exp \left(-\tau \frac{\hat{V}}{2} \right)
\exp \left(-\tau \hat{H}_0 \right) \times \nonumber \\
\exp \left(-\tau    \frac{\hat{V}}{2} \right)+
\cdots
  \label{fractorization2}
\end{eqnarray}
and
\begin{eqnarray}
    \exp \left[-\tau(\hat{H}_0+\hat{V}) \right]=
\exp \left(-\tau \frac{\hat{V}}{6} \right)
    \exp \left(-\tau \frac{\hat{H}_0}{2} \right) \times \nonumber \\\
\exp \left(-\tau \frac{2\tilde{{V}}}{3} \right)
\exp \left(-\tau \frac{\hat{H}_0}{2} \right)
\exp \left(-\tau \frac{\hat{V}}{6} \right) \nonumber \\
+ \cdots ,
  \label{fractorization4}
\end{eqnarray}
where $\tilde{V}$ is 
given by $\hat{V}+\tau^2[\hat{V},[\hat{H}_0,\hat{V}]]/48$.
For observables that are determined by the
diagonal elements of the density matrix,
these factorizations yield errors that scale as $\tau^3$ and $\tau^5$,
respectively~\cite{chin97}. 
The non-interacting part of the density matrix 
in the position basis can be written
compactly~\cite{pathria1996statistical,krauth2006statistical},
\begin{eqnarray}
  \langle \mathbf{R}|e^{-\tau
  \hat{H}_0}|\mathbf{R}' \rangle=a_{\text{ho}}^{-3N}\left[2\pi\sinh({\tilde{\beta}})\right]^{-3N/2}   \times  \nonumber \\
  \exp \left[
-\frac{(\mathbf{R}^2+\mathbf{R}'^2)
\cosh(\tilde{\beta})-
2\mathbf{R}\cdot\mathbf{R}'}{2\sinh(\tilde{\beta}) a_{\text{ho}}^{2}}
\right].
 \label{NonInteractingTrap}
\end{eqnarray}
Here, $\tilde{\beta}$
denotes the dimensionless 
inverse temperature, $\tilde{\beta}=\beta E_{\text{ho}}$.
The potential dependent part of the density matrix reduces to evaluating
the potential
at the given configuration.

The energy and structural expectation values are
calculated following standard procedures~\cite{CeperleyRMP}.
The superfluid fraction is calculated using the area 
estimator~\cite{ceperley87,ceperley89,CeperleyRMP}.
The superfluid density is calculated following Ref.~\cite{kwon06}.
The condensate fraction requires off-diagonal elements
of the density matrix, i.e., open paths~\cite{ceperleycondensate86}. 
We have not yet implemented 
this.

In the high temperature limit, the particle statistics becomes negligible and
the system behaves, to leading order, as a non-interacting gas of Boltzmann
particles.
To analyze the effects of the particle statistics
for systems with two or more identical particles
in the low temperature regime, we find
it useful to divide the partition function
$Z$ into ``even'' and ``odd''
contributions (a closely related definition can be found in
Ref.~\cite{FermiSignProblem}),
\begin{equation}
  Z=Z_{\text{even}} \pm Z_{\text{odd}},
\end{equation}
where
\begin{eqnarray}
Z_{\text{even}}=
\sum_{P_{\text{even}}}\int \mathrm{d}\mathbf{R}\rho(\hat{\mathcal{P}}\mathbf{R},\mathbf{R},\beta)
\end{eqnarray}
and
\begin{equation}
Z_{\text{odd}}=
  \sum_{P_{\text{odd}}}\int \mathrm{d}\mathbf{R}\rho(\hat{\mathcal{P}}\mathbf{R},\mathbf{R},\beta);
  \label{<++>}
\end{equation}
the plus and minus sign apply if the system contains identical
bosons and fermions, respectively (here and in the remainder of this
section we assume that the system contains only one type of
identical particles).
The
sum over $P_{\text{even}}$ includes the permutations that are
characterized by even $N_{\text{I}}(\sigma)$ 
and the sum over $P_{\text{odd}}$ includes the
permutations that are
characterized by odd $N_{\text{I}}(\sigma)$.
The sum over $P_{\text{odd}}$ is only non-zero if the system
under study contains two or more identical particles.
When the temperature is high, only the 
identity permutation (and thus only the first term) contributes, i.e.,
the 
statistics is suppressed and the system behaves
like a Boltzmann gas.
As the temperature decreases, the relative importance of the second term
increases.
In the zero temperature limit, the two terms contribute equally.
We define the 
statistical factor $S$ as the normalized ratio of the 
``even'' and ``odd'' partition
functions~\cite{FermiSignProblem}~\footnote{For
 identical fermions, our $S$ is
 equal to the quantity $\unexpanded{\langle S\rangle}_{\tilde P}$ of
 Ref.~\protect\cite{FermiSignProblem}},
\begin{equation}
S=
\frac{Z_{\text{even}}-Z_{\text{odd}}}{Z_{\text{even}} + Z_{\text{odd}}}.
  \label{<++>}
\end{equation}
The statistical factor $S$ approaches 
$1$ in the high-temperature limit and $0$ in the zero-temperature limit.
Since the partition function enters into 
the denominator of the thermal expectation values, 
the statistical factor characterizes
the 
numerical demands on the simulation for systems with
identical fermions.
The smaller $S$ is, the harder the simulation is. 
As a rule of thumb,
if we compare the $S$ value for the same system at two
different temperatures, then the simulation time required to
obtain comparable accuracy for the observables at the
two temperatures 
is $(S_{\text{high}}/S_{\text{low}})^2$
times larger at the lower temperature than at 
the higher temperature (here, $S_{\text{high}}$ and 
$S_{\text{low}}$ are the $S$
values at the higher and lower temperature, respectively).
 This phenomenon
is known as the Fermi sign
problem~\cite{stratonovich1957method,ceperley1986quantum,ceperley1991fermion,FermiSignProblem}.
A related interpretation of $S$ is in terms of
the ``quantum statistics'' of the system under study.
For both bosons and fermions,
a value of $S$ around $1$ indicates that the particles approximately follow
Boltzmann statistics while a value of $S$ close to $0$
indicates that exchange effects play an important role.

\begin{figure}
\centering
\includegraphics[angle=0,width=0.4\textwidth]{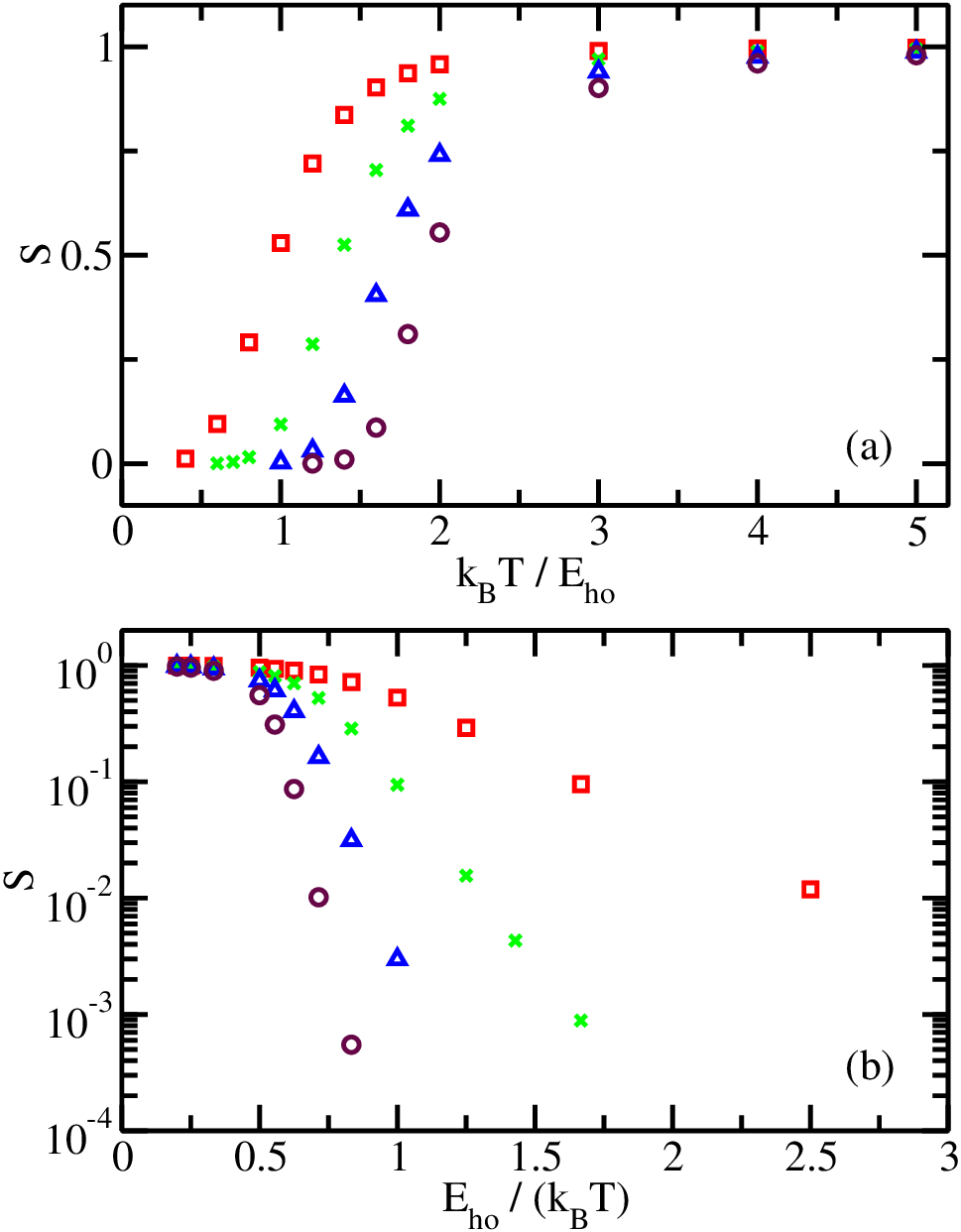}
\caption{(Color online)
  Statistical factor $S$ for the $(N-1,1)$ system with interspecies potential
  $V_{\text{G}}$ with $r_0=0.06a_{\text{ho}}$ and $1/a_s=0$. 
  Squares, crosses, triangles, and circles 
  show the statistical factor $S$
  as a
  function of (a) the temperature $T$ and (b) the inverse temperature
  $T^{-1}$
  for $N=3, 4,5$, and $6$, respectively.
 }\label{statisticalfactor}
\end{figure}

Figure~\ref{statisticalfactor}(a) 
shows the statistical factors as a function of
the temperature for the
$N$ particle system consisting of $N-1$ 
identical particles and one impurity.
The identical particles do not interact while
the unlike particles interact through a
Gaussian potential with $r_0=0.06a_{\text{ho}}$ and infinite $s$-wave
scattering length.
The statistical factor deviates
notably from one when the temperature is of the order of
the ``Fermi temperature'' or lower.
The Fermi temperature is equal to $5 E_{\text{ho}}/2$ to
$7 E_{\text{ho}}/2$
for the $(N-1,1)$ systems with $N=3-6$.
At low temperature, the statistical factor depends exponentially
on the inverse temperature, i.e., 
$S\propto\exp(-\beta \alpha_N)$~\cite{FermiSignProblem},
where
$\alpha_N$ increases faster than linear with
increasing $N$.
We have performed reliable
calculations for the symbols shown in
Fig.~\ref{statisticalfactor}. The
lowest temperature that can be reached depends,
of course, on the available computational resources. However, since
the Fermi sign problem increases exponentially with decreasing
temperature, the lower $T$ limit shown in Fig.~\ref{statisticalfactor}
is somewhat generic.
The physics of the $(N-1,1)$ systems 
with Bose, Fermi and Boltzmann statistics is discussed in more detail 
in Sec.~\ref{sec_mixture}.

\section{Condensate and superfluid fractions of the two-body system}
\label{sec_twobody}
The
condensate and superfluid fractions are distinct physical
quantities that vanish when the de Broglie wave length
is small but differ from zero when the de Broglie wave length is
large.
This section compares the condensate and superfluid fractions
for the simplest interacting system,
namely for two particles in a harmonic trap with zero-range $s$-wave
interactions.
For this system,
the eigen spectrum and eigen functions are known
in compact analytical form~\cite{busch}, which facilitates the 
calculation of $n_c$ and $n_s$
over a wide temperature range.
The superfluid fraction is
calculated using the energy eigen states
in the moment of inertia based definition
[see Eq.~(\ref{eq_superfluidfraction})].

An important point of this section is that the 
superfluid and condensate fractions are
meaningful quantities not just for large systems but also for small systems.
We will show in Sec.~\ref{sec_boson} that the superfluid fraction
of the $N$ boson system is, for certain parameter combinations,
approximated well by that of a single particle.
The superfluid fraction reflects symmetry properties of
the system~\cite{persistentcurrents,farrelly14,OurArxivPaper}. 
The connection
between superfluidity and angular momentum decoupling mechanisms,
e.g.,
has been discussed in some detail in the context of
small doped bosonic helium droplets~\cite{JCPPaperByKwonHuangBlumeWhaley,WhaleyandFarrelly}.
The condensate fraction is given by the largest
eigen value of the
one-body reduced density matrix $\rho_{\text{red}}$
or, equivalently, the largest occupation number of
the
natural orbitals~\cite{famousWorkByOnsager,leggettbook,LookAtTheReferencesCitedByCederbaumsGroup}. Since
the natural orbitals are defined by decomposing the reduced
density matrix in a specific way, the occupation numbers,
and hence the condensate fraction, can be interpreted as 
a particular measure of the particle-particle correlations of the
system.
Our approach for determining the finite temperature
reduced density
matrix of the two-body system (which is discussed in the following
paragraphs) also allows one to
determine entanglement 
measures such as 
the concurrence~\cite{concurrence} and negativity~\cite{negativity} 
of the two-particle system over a 
wide temperature range. Such calculations appear to have been challenging
in the past~\cite{prayou06}.

The reduced density matrix $\rho_{\text{red}}$
for the two-particle system reads
\begin{eqnarray}
\label{eq_condfrac1}
\rho_{\text{red}}(\mathbf{r}_1',\mathbf{r}_1,\beta)=
Z^{-1}
\int d\mathbf{r}_2 
\rho(\mathbf{r}_1',\mathbf{r}_2,\mathbf{r}_1,\mathbf{r}_2,\beta).
\end{eqnarray}
Using the separation of the center of mass and relative coordinates,
Eq.~(\ref{eq_condfrac1}) becomes
\begin{eqnarray}
\label{eq_condfrac2}
\rho_{\text{red}}(\mathbf{r}_1',\mathbf{r}_1,\beta)= \nonumber \\
Z^{-1} \int d\mathbf{r}_2 
\rho_{\text{rel}}(\mathbf{r}_{\text{rel}}',\mathbf{r}_{\text{rel}},\beta)
\rho_{\text{cm}}(\mathbf{r}_{\text{cm}}',\mathbf{r}_{\text{cm}},\beta),
\end{eqnarray}
where 
$\mathbf{r}_{\text{rel}}=\mathbf{r}_1-\mathbf{r}_2$,
$\mathbf{r}_{\text{rel}}'=\mathbf{r}_1'-\mathbf{r}_2$,
$2\mathbf{r}_{\text{cm}}=\mathbf{r}_1+\mathbf{r}_2$, 
$2\mathbf{r}_{\text{cm}}'=\mathbf{r}_1'+\mathbf{r}_2$,
\begin{eqnarray}
\label{eq_denmatrel}
  \rho_{\text{rel}}(\mathbf{r}_{\text{rel}}',\mathbf{r}_{\text{rel}},\beta)
= \nonumber \\
\sum_{ilm}e^{-\beta
    E_{i,l}}\psi_{ilm}^*(\mathbf{r}_{\text{rel}}')
\psi_{ilm}(\mathbf{r}_{\text{rel}})  
,
\end{eqnarray}
and
\begin{eqnarray}
  \label{eq_denmatcm}
  \rho_{\text{cm}}(\mathbf{r}_{\text{cm}}',\mathbf{r}_{\text{cm}},\beta)
= \nonumber \\
    \sum_{QLM}e^{-\beta
      E_{Q,L}}\psi_{QLM}^*(\mathbf{r}_{\text{cm}}')
\psi_{QLM}(\mathbf{r}_{\text{cm}})      .
\end{eqnarray}
In Eq.~(\ref{eq_denmatcm}),
$E_{Q,L}$ denotes the center-of-mass eigen energy,
which can be conveniently written in terms
of the principal quantum number $Q$ ($Q=0,1,\cdots$) and
the center of mass angular momentum quantum
number $L$ ($L=0,1,\cdots$),
$E_{Q,L}=(2Q+L+3/2) E_{\text{ho}}$.
The energies are
independent of the projection quantum number $M$ ($M=-L,-L+1,\cdots,L$).
In Eq.~(\ref{eq_denmatrel}),
$E_{i,l}$ denotes the relative eigen energy.
For two Boltzmann particles, all $l$ values are allowed.
For two identical bosons, in contrast, only even $l$ values
are allowed.
For finite relative angular momentum $l$,
the relative energy reads $E_{i,l}=(2 i + l +3/2) E_{\text{ho}}$,
where $i=0,1,\cdots$.
For $l=0$, $i$ denotes a non-integer quantum number whose values
are determined semi-analytically by solving a transcendental 
equation~\cite{busch}.
As in the center of mass case, the relative energies are independent of
the projection quantum number $m$ ($m=-l,-l+1,\cdots,l$).

To evaluate $\rho_{\text{rel}}$, we use the fact that the $l>0$
states are not affected by the zero-range interactions
and write
$\rho_{\text{rel}}=
\rho_{\text{rel}}^{l>0,\text{NI}}+
\rho_{\text{rel}}^{l=0,\text{int}}$,
where 
$\rho_{\text{rel}}^{l>0,\text{NI}}$
denotes the $l>0$ contributions to the density matrix 
(these contributions are independent of the $s$-wave scattering length)
and 
$\rho_{\text{rel}}^{l=0,\text{int}}$ the $l=0$ contribution that
depends on $a_s$.
To evaluate the latter,
it is convenient to project the interacting $l=0$ 
energy eigen states onto the non-interacting 
harmonic oscillator states,
$\psi_{i00}(\mathbf{r}_{\text{rel}})=
\sum_{q=0}^\infty C_q^{(i)} \psi_{q00}(\mathbf{r}_{\text{rel}})$,
where $q=0,1,\cdots$.
The expansion coefficients $C_q^{(i)}$ are known 
analytically~\cite{busch,KevinDesmondPRA}.
Now that $\rho_{\text{cm}}$ and $\rho_{\text{rel}}$ are expressed in terms of
the non-interacting wave functions
in the relative and center of mass coordinates, the integral
over $d\mathbf{r}_2$ can be performed
by reexpressing, using the
Talmi-Moshinsky brackets~\cite{talmi52,moshinsky59},
 the harmonic oscillator eigen states in the relative
and center of mass coordinates in terms of the 
harmonic oscillator eigen states in the single particle
coordinates.
After integrating over $d \mathbf{r}_2$, we project the
reduced density matrix onto single-particle states in the $\mathbf{r}_1$
coordinate.
Using the orthogonality
of the Clebsch-Gordon coefficients
as well as other standard identities 
from angular momentum algebra, 
the calculation of the matrix elements simplifies
dramatically.
The resulting one-body density matrix is
found to be block
diagonal in the $l$ and $m$ quantum numbers.
Furthermore, since the lowest $l=0$
state always minimizes
the energy, 
the largest occupation number
comes from the $(l,m)=(0,0)$ submatrix.
The results discussed in the following are obtained by
diagonalizing a $20 \times 20$ submatrix.
Increasing the matrix size to $50 \times 50$
changes the results by less than $1\%$.

\begin{figure}
\centering
\includegraphics[angle=0,width=0.4\textwidth]{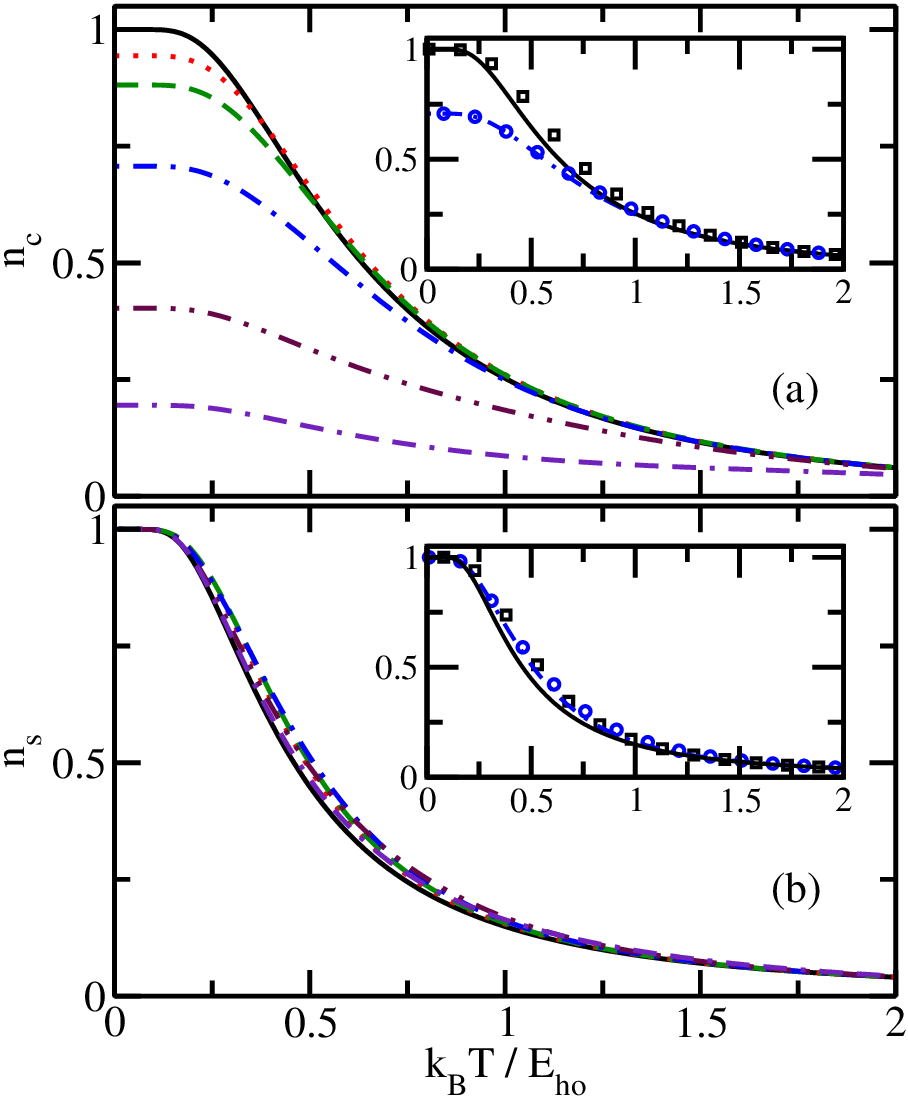}
\caption{(Color online)
The lines show
  (a) the condensate fraction $n_c$ 
and (b) the superfluid fraction $n_s$ 
as a function of the temperature $T$ for two
  Boltzmann particles with zero-range interaction for various $a_s$.
  The solid, dotted, dashed,
  dash-dotted, dash-dot-dotted, and dash-dash-dotted lines are for
  $a_{\text{ho}}/a_{s}=-\infty, -2, 1, 0, 1,$ and $2$, respectively.
  In panel (b), the dependence on $a_s$ is small.
The insets compare (a) the condensate fraction $n_c$
and (b) the superfluid fraction $n_s$ for
two Boltzmann particles (lines; these are the same data as shown
in the main parts of the figure) and
two identical bosons (squares and circles
correspond to $a_{\text{ho}}/a_s=-\infty$ and $0$,
respectively) as a function of the temperature.
 }\label{11con}
\end{figure}

The main panel of Fig.~\ref{11con}(a) shows the condensate fraction 
$n_c$ for two Boltzmann
particles as a function of the 
temperature for various $s$-wave scattering lengths $a_s$. 
Solid, dotted, dashed, dash-dotted, dash-dot-dotted
and dash-dash-dotted lines are for 
$a_{\text{ho}}/a_{s}=-\infty, -2, -1, 0, 1,$ and
$2$, respectively. 
As the temperature increases, the condensate fraction $n_c$ decreases for
all interaction strengths. 
At zero temperature, $n_c$ decreases as the inverse
scattering length increases.
At finite temperature,
however, we observe in some cases 
[see the $k_B T \approx E_{\text{ho}}/2$ to
$E_{\text{ho}}$ regime in Fig.~\ref{11con}(a)] that
the condensate fraction increases slightly as $|a_s|$ ($a_s<0$) increases.
This is caused by the interplay of the interaction energy and
the temperature dependent Boltzmann weight.

The condensate fraction for two identical bosons is very similar to
that for two Boltzmann particles.
The inset of Fig.~\ref{11con}(a) compares the condensate fraction
for two identical bosons (symbols) with those for two
Boltzmann particles (lines) for $a_{\text{ho}}/a_s= -\infty$ and
$0$, respectively. 
It can be seen that the condensate fraction for 
two identical bosons falls off slightly slower with increasing 
temperature than that for two Boltzmann particles.
This is because the Bose statistics excludes the states with
odd relative 
angular momentum $l$, implying that the $l=0$
states (which are responsible
for the non-zero
condensate fraction) 
are relatively more important for two identical bosons
than for two Boltzmann particles.

For comparison, Fig.~\ref{11con}(b) shows the superfluid fraction 
$n_s$ for two Boltzmann
particles for the same scattering lengths. 
The superfluid fraction $n_s$ depends
weakly on the $s$-wave scattering length.
Specifically, the superfluid fraction approaches
$1$ in the low temperature regime for all
$s$-wave scattering lengths.
This is a consequence of the fact that the lowest energy eigen state
has vanishing total orbital angular momentum for all
$s$-wave scattering lengths. The inset of Fig.~\ref{11con}(b)
compares the superfluid fraction for two Boltzmann particles (lines)
with those for two identical bosons (symbols).
As in the case of the condensate fraction, the switch
from Boltzmann to Bose statistics changes the
superfluid fraction only by a small amount.

A comparison of Figs.~\ref{11con}(a) and \ref{11con}(b)
shows that the condensate and superfluid fractions are distinctly different
quantities.
When the two-body system forms a molecule 
(for positive $a_s$),
the condensate fraction is small.
The superfluid fraction, in contrast, remains 
approximately 1 in the low temperature regime,
indicating that the response to an infinitesimal rotation
is largely independent of 
the size of the system (the density decreases
with increasing $1/a_s$) and instead largely determined by its
spherical shape.

\begin{figure}
\centering
\includegraphics[angle=0,width=0.4\textwidth]{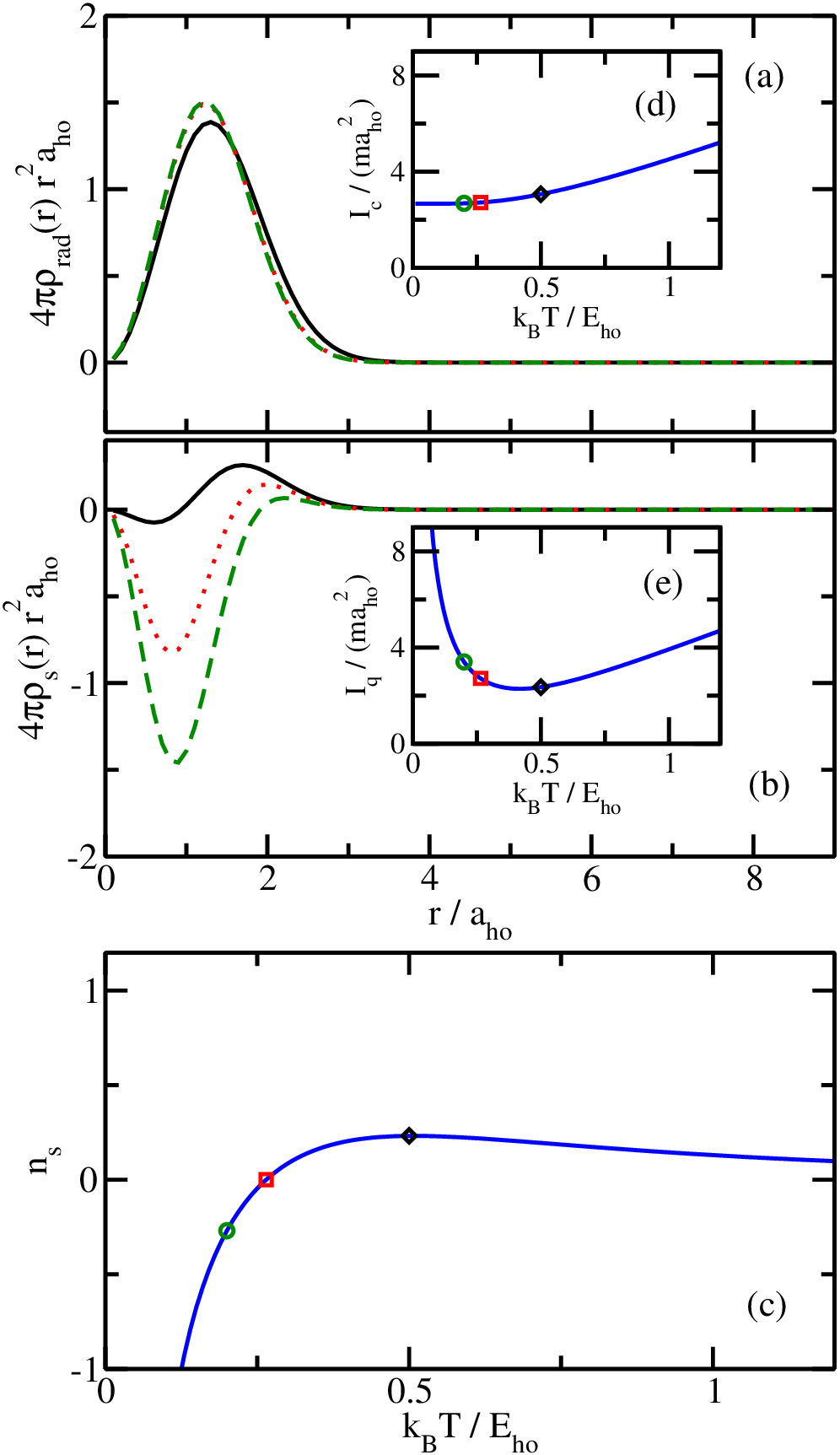}
\caption{(Color online)
Panels (a) and (b) show radial densities for two identical non-interacting fermions.
  Solid, dotted, and dashed lines show
  (a) the scaled radial total density and
  (b) the scaled radial superfluid density,
  for $k_BT/E_{\text{ho}}
  = 0.5, 0.26459,$ and $0.2$, respectively.
  In panel (a), the dotted line is hardly distinguishable from the dashed
  line.
The solid lines in panels (c), (d), and (e) show (c) the superfluid fraction $n_s$, 
(d) the classical moment
of inertia $I_{\text{c}}$, and (e) the quantum mechanical 
moment of inertia $I_{\text{q}}$ 
as a function
of the temperature $T$.
The diamond, square and circle mark the temperatures considered
in panels (a) and (b).
 }\label{2fermions}
\end{figure}

Next, we consider two identical fermions.
Naively, this system might be thought to be ``uninteresting''
since the Pauli exclusion principle prohibits
scattering in the $s$-wave channel. As we show now, two non-interacting
identical fermions display intriguing temperature-dependent
behaviors.
For two identical non-interacting fermions, the condensate fraction
equals $1/2$ at $T=0$ and decreases monotonically.
The superfluid fraction displays [see Fig.~\ref{2fermions}(c)]
a non-monotonic dependence on the temperature. As expected,
$n_s$ is zero in the high $T$ limit, increases
to around $0.2$ at $k_BT= E_{\text{ho}} /2$, and then diverges to 
$-\infty$ in the zero temperature limit. As discussed in 
Ref.~\cite{OurArxivPaper},
this behavior can be understood by analyzing the 
classical moment of inertia $I_{\text{c}}$ and the quantum moment
of inertia $I_{\text{q}}$ [see the
Figs.~\ref{2fermions}(d) and \ref{2fermions}(e), respectively].
Specifically, the fact that the lowest energy eigen state has 
$L_{\text{tot}}=1$ is responsible for the increase of $I_{\text{q}}$ at low
temperature. Motivated by the nuclear physics
literature~\cite{migdal59,liu05}, 
we refer to this behavior
as ``abnormal''.

The fact that the superfluid fraction for two identical
fermions becomes negative in the
low-temperature regime can be understood as follows~\cite{OurArxivPaper}.
Two identical bosons at low temperature do not respond to
an infinitesimal external rotation ($n_s \rightarrow 1$ as
$T \rightarrow 0$) 
since the lowest energy eigen state
has $L_{\text{tot}}=0$.
Two identical fermions at low temperature, however, do respond to
an infinitesimal external rotation 
($n_s \rightarrow -\infty$ as
$T \rightarrow 0$) 
since the lowest energy eigen state
has $L_{\text{tot}}=1$. The physical picture is
that the system ``speeds
up'' faster than we would expect for a normal fluid with the same
classical moment of inertia~\cite{OurArxivPaper}.

To gain further insight into the superfluid properties
of the fermionic system, we analyze the radial and superfluid densities.
The radial densities for particles $1$ and $2$ are identical and the subscript $j$
of $r_j$ will be dropped in what follows.
Solid, dotted and dashed
lines in
Fig.~\ref{2fermions}(a) 
show the scaled radial density $4 \pi \rho_{\text{rad}}(r)r^2$
for $k_BT/E_{\text{ho}}=0.5$,
$0.26459$, and $0.2$, respectively.
The radial density is fairly insensitive to the temperature.
The radial superfluid density [see Fig.~\ref{2fermions}(b)], in contrast,
changes notably with the temperature.
This is not unexpected since the superfluid fraction
varies strongly in the low temperature regime.
The radial superfluid density takes negative values near the trap center and 
positive values near the edge of the cloud. The oscillation
of the radial superfluid density reflects the fact that the
lowest energy eigen state has total angular momentum quantum
number $L_{\text{tot}}=1$.
For large $r$, the probability of finding two particles close to
each other is extremely low.
This translates into the Fermi statistics playing a negligible role.
On the other hand, we expect that the Fermi statistics is much more
important near the trap center.
In the language of path integrals,
the ``permuted  paths'' 
(i.e., the paths that come from exchanging 
particles 1 and 2 and thus contribute with a negative 
sign to the partition function)
are largely concentrated near the center. 
These  ``permuted paths'' contribute negatively to the area estimator and
span larger areas compared to the ``unpermuted paths''.
As a consequence,
the superfluid density is negative near the trap center.

The analysis presented here for two non-interacting identical fermions
can be extended to two-component Fermi gases 
with interspecies $s$-wave interactions consisting of $N=3$ or more particles.
Selected results were presented in our earlier work~\cite{OurArxivPaper}.
We anticipate that the analysis
of the superfluid properties presented in the previous 
paragraphs for two non-interacting fermions will inspire
other studies, for bosons or fermions,
that are concerned with understanding the distribution of
the superfluid properties in finite sized systems or
systems with interfaces~\cite{vycor88,JCPPaperByKwonHuangBlumeWhaley,kwon02,kwon05,kwon06,boninsegnipra07,boninsegniprl08,prokofev05,prokofev07}.

\section{$N$-body systems}
\label{sec_nbody}
\subsection{$N$ identical bosons}
\label{sec_boson}
\begin{figure}
\centering
\includegraphics[angle=0,width=0.4\textwidth]{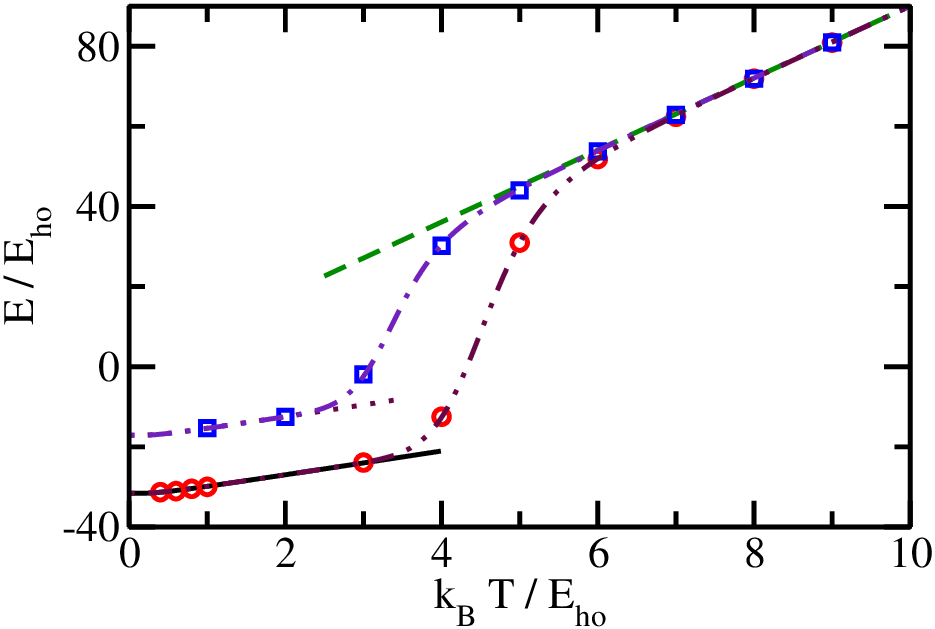}
\caption{(Color online)
  Energies as a function of the temperature $T$ for 
three identical bosons at unitarity
  interacting through $V_{\text{G}}$ with different $r_0$.
  Circles and squares show the PIMC results 
  for $r_0/a_{\text{ho}}=0.06$ and $0.08$,
  respectively.
  For comparison, the solid and dotted lines show the result
  obtained using the droplet state plus center of mass excitations.
  The dashed line shows the thermally averaged
  energy for three identical non-interacting 
  bosons.
  Dash-dot-dotted and dash-dotted lines show results obtained using the
  simple combined model for $r_0/a_{\text{ho}}=0.06$ and 
$0.08$ (see the text for discussion).
 }\label{3benergy}
\end{figure}

This section discusses the temperature dependent properties
of $N$ identical bosons under external spherically symmetric harmonic
confinement interacting through the 
Gaussian model potential $V_{\text{G}}$ with infinite $s$-wave
scattering length.
Circles and squares in Fig.~\ref{3benergy}
show the energy of the three-boson
system, obtained from the PIMC simulations, 
as a function of the temperature for $r_0/a_{\text{ho}}=0.06$ and $0.08$,
respectively.
For both ranges, the energy shows three distinct regions.
The energy increases approximately linearly at small $T$, turns up
relatively sharply around $k_BT= 4E_{\text{ho}}$ or $3 E_{\text{ho}}$,
and then changes again linearly.
The energy at low temperature---if expressed in harmonic oscillator units---shows a strong range dependence.
The energy at high temperature, in contrast, is to leading order
independent of $r_0$.
We refer to the rapid change of the energy from one approximately
linear regime to the other approximately linear regime as a
phase transition
like feature.

We now introduce a simple parameter-free model that 
reproduces the energy curves semi-quantitatively 
(see the dash-dot-dotted and dash-dotted lines in Fig.~\ref{3benergy}).
The assumptions going into the model are
that the low-temperature behavior
is governed by the properties of the lowest Efimov trimer
and that the high-temperature
behavior is governed by the properties 
of the non-interacting three-boson gas.
Treating only the lowest Efimov trimer state and its
center of mass excitations, we obtain the solid and dotted lines
in Fig.~\ref{3benergy} for $r_0/a_{\text{ho}}=0.06$ and $0.08$,
respectively.
These thermally averaged energies are obtained using the lowest
eigen energy
of the trapped three-boson system, i.e., using the
eigen energy of the state that shows Efimov characteristics, and summing
over the center of mass excitations. 
The dashed line shows the thermally averaged
energy of three non-interacting identical bosons.
If we combine these two limiting behaviors,
the model partition function $Z_{\text{model}}$ reads
\begin{eqnarray}
\label{eq_combinedmodel}
Z_{\text{model}}(\beta) = Z_{\text{droplet}}(\beta) + Z_{\text{gas}}(\beta),
\end{eqnarray}
where
$Z_{\text{droplet}}(\beta)=z(\beta)\exp(-\beta E_{\text{droplet}})$
and
$Z_{\text{gas}}(\beta) =[z^3(\beta)+3z(2\beta)z(\beta)+2z(3\beta)]/6$.
Here, $E_{\text{droplet}}$ denotes the lowest relative 
eigen energy of the three-boson system and
$z(\beta)$ the partition function of a single
harmonically trapped particle.
The second and third terms in
$Z_{\text{gas}}$ originate from the symmetrization of $Z_{\text{gas}}$.
The resulting energies are shown 
in Fig.~\ref{3benergy} by the dash-dot-dotted and dash-dotted
lines for
$r_0/a_{\text{ho}}=0.06$ and $0.08$,
respectively.
The agreement between this simple combined model and the PIMC calculations
is very good.

One may ask why the simple combined model works so well.
We attribute this to primarily two things. First, 
for the examples shown in 
Fig.~\ref{3benergy} the energy separation
between the lowest Efimov trimer state and the gas-like states is large
(the case where $|E_{\text{droplet}}|$ is not much
larger than $E_{\text{ho}}$ is briefly discussed at the
end of this section).
Second, although the system is strongly-interacting,
the non-interacting Bose gas model describes the density
of states approximately correctly.
The reason is that a significant fraction of the states is not
affected by the $s$-wave interactions~\cite{WernerandCastinPRL}.
In fact, if we replace
the partition function 
$Z_{\text{gas}}$ for the non-interacting Bose gas by the
partition function for the non-interacting Boltzmann gas,
then the model predicts that the energy changes rapidly at a 
lower temperature than predicted by the PIMC results.
If, on the other hand,
we replace
the partition function 
$Z_{\text{gas}}$ for the non-interacting Bose gas by a partition function for three identical bosons 
that accounts for the $s$-wave interactions in an approximate manner
(we reduce the energy of all states that are affected by the $s$-wave interactions
by $E_{\text{ho}}$), the resulting energy curves are, on the scale of
Fig.~\ref{3benergy}, 
indistinguishable from the dash-dot-dotted and dash-dotted
curves.

\begin{figure}
\centering
\includegraphics[angle=0,width=0.45\textwidth]{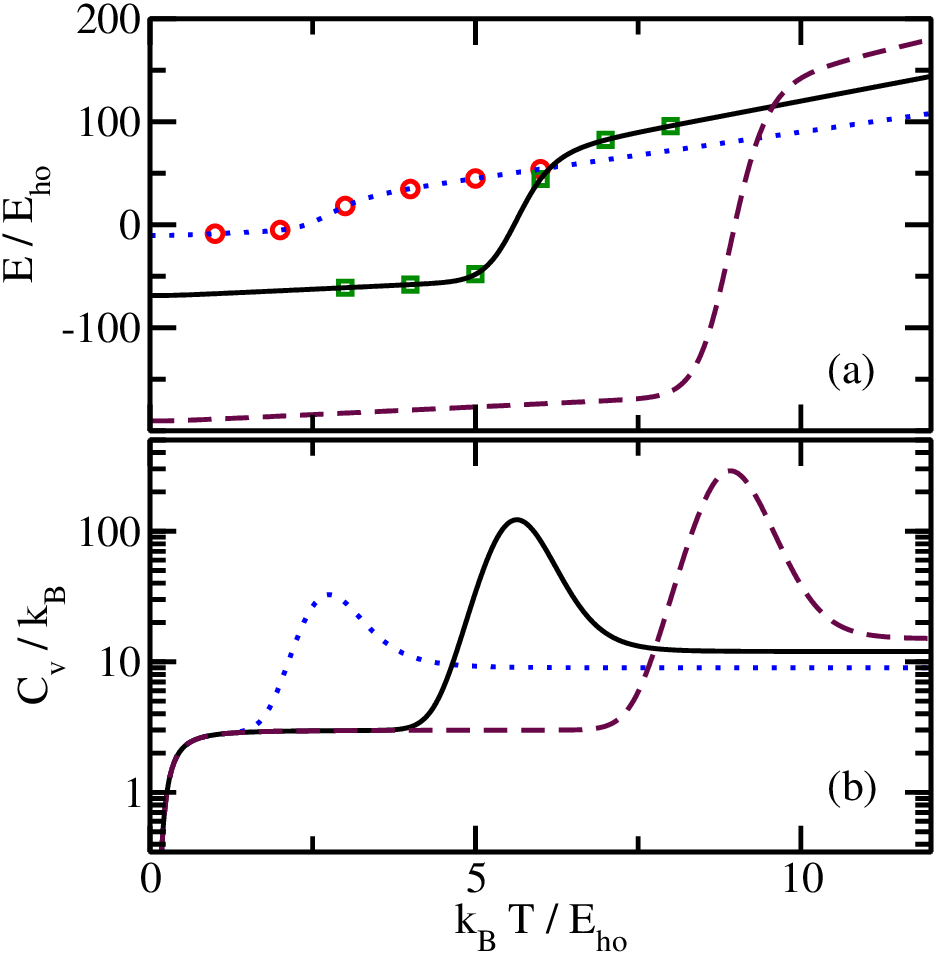}
\caption{(Color online)
  Phase transition like feature for $N$ identical harmonically trapped bosons
  interacting through $V_{\text{G}}$ 
with $1/a_s=0$. (a) Circles and squares show the
  energy obtained by the PIMC approach for $r_0=0.1a_{\text{ho}}$ 
and $N=3$ and $4$,
  respectively, as a function of the temperature $T$.
  The dotted, solid, and dashed lines show the energies for $N=3$, $4$, and
  $5$ obtained using the simple combined model. (b) The dotted, solid and
  dashed lines show the heat capacity $C_v$ for $N=3$, $4$, and $5$,
  respectively, as a function of $T$.
 }\label{345energy}
\end{figure}

Circles and squares in Fig.~\ref{345energy}(a)
show the thermally averaged PIMC energies for the Gaussian model
interaction with $r_0/a_{\text{ho}}=0.1$ and $1/a_s=0$ for
$N=3$ and $4$, respectively.
As the three-boson system,
the four-boson system displays a ``phase transition like'' feature.
To model four- and higher-body boson systems, we generalize the combined model
introduced above as follows.
In Eq.~(\ref{eq_combinedmodel}),
$Z_{\text{droplet}}(\beta)$ now denotes the 
partition function determined by the lowest $N$-boson
energy state plus center of mass excitations
and $Z_{\text{gas}}(\beta)$ denotes the
partition function of the non-interacting $N$-boson gas.
As above, $Z_{\text{gas}}$ is properly symmetrized.
The solid line in Fig.~\ref{345energy}(a) shows the resulting energy
for the four-boson system. The agreement with the 
PIMC results is good.
It should be noted that the combined model neglects,
for systems with $N>3$, a 
large number of states.
For example, for the four-boson system, it neglects the excited 
four-boson Efimov state whose energy is,
in the universal regime, $1.002$ times 
the trimer energy~\cite{deltuvaFewBodySystemsPaper} 
as well as ``atom-trimer states'' that can be
approximately described as consisting of 
an Efimov trimer with the fourth particle occupying one of the
harmonic oscillator states.
These states contribute relatively little to the partition function
for two reasons. First, the separation between the four-body ground 
state energy
and the energy of the excited tetramer and the separation
between the four-body ground state energy and the atom-trimer states
is large (the factor for the former is $4.61$ in the universal 
regime~\cite{deltuvaFewBodySystemsPaper}).
Second, the density of states of the atom-trimer states is negligible 
compared to the
density of states of the gas-like boson-boson-boson-boson states.
We conjecture that the combined model also provides a good description
for larger Bose systems.
We stress that
the combined model
is fully analytical, provided that the eigen energy of the lowest 
$N$-body state, which can be considered as being
tied to the lowest trimer eigen state, is known.
The dashed line in Fig.~\ref{345energy}(a) shows the 
energy for
$N=5$ bosons interacting through $V_{\text{G}}$
with $r_0/a_{\text{ho}}=0.1$ and $1/a_s=0$ as a function of 
the temperature.
This curve is
obtained using the combined model with the eigen energy of the lowest
$N=5$ energy eigen state as input (see Table~\ref{energytable} for the
energy).

Figure~\ref{345energy}(a) shows 
that the phase transition like feature
for fixed $r_0$
moves to higher temperature with increasing $N$.
To estimate the transition temperature $T_{\text{tr}}$,
we calculate the heat capacity $C_v$,
$C_v= \partial E / \partial T$.
The dotted, solid and dashed lines in 
Fig.~\ref{345energy}(b) show $C_v$,
obtained using the combined model for the thermally
averaged energy [see lines in Fig.~\ref{345energy}(a)], 
as a function of the temperature
for $N=3,$ $4$, and $5$, respectively. 
The heat capacity curves show distinct maxima. 
We define the transition temperature $T_{\text{tr}}$ 
as the temperature at which the heat capacity takes 
on its maximum.

\begin{figure}
\centering
\includegraphics[angle=0,width=0.45\textwidth]{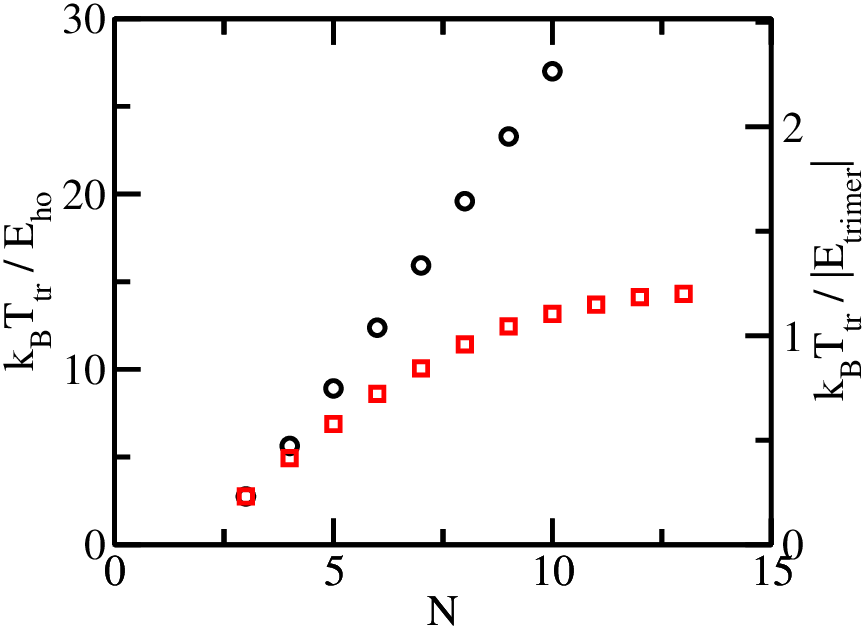}
\caption{(Color online)
  Transition temperature $T_{\text{tr}}$ for $N$ identical bosons in a
  harmonic trap at unitarity
  as a function of $N$.
  The transition temperature is calculated using the
  simple combined model.
  The circles show $T_{\text{tr}}$ 
  using the droplet energies for the Gaussian two-body interaction model
  employed in this work. For comparison, the squares show 
  $T_{\text{tr}}$ using the droplet energies for a model
  Hamiltonian with attractive two-body and repulsive three-body
  interactions~\cite{vonStecherJPB} (to obtain the squares,
the three-body eigen energy
  $E_{\text{droplet}}=E_{\text{trimer}}$ 
  is chosen such that it agrees with that for the Gaussian two-body 
interaction model, i.e., the circle and the square
agree for $N=3$).
 }\label{transition}
\end{figure}

The circles in Fig.~\ref{transition}
show the transition temperature 
for $N$ bosons interacting through
$V_{\text{G}}$ with $r_0/a_{\text{ho}}=0.1$ and 
$1/a_s=0$ as a function of $N$.
To obtain the transition temperature, we extrapolate the
PIMC energies at low temperature to the zero
temperature limit.
The resulting zero-temperature energies $E_{\text{droplet}}$
are reported in Table~\ref{energytable}.
\begin{table}
  \centering
  \caption{Relative zero-temperature energy $E_{\text{droplet}}$  for $N$ bosons
  interacting through the Gaussian potential $V_{\text{G}}$ with diverging $s$-wave scattering length.
  The energies in columns 2 and 5 are obtained by extrapolating the PIMC
  results to $T=0$. The energies are expressed in units of the short-range
  energy scale $E_{\text{sr}}$, $E_{\text{sr}}=\hbar^2/(mr_0^2)$.
  Column 3 reports the energies from Ref.~\cite{kievskyarxiv}; no errorbars are
  reported in that reference.
  For comparison, our basis set expansion approach (see Ref.~\cite{Rakshit}
  for a discussion of the approach) yields
  $E_{\text{droplet}}/E_{\text{sr}}=- 0.11923(1)$ and $-0.70173(5)$ for $N=3$ and $4$,
  respectively.
}
\begin{tabular} 
{c d{4.7} d{4.4} |c d{4.7}}
    \hline
    \hline
    $N$ & \multicolumn{1}{c}{$E_{\text{droplet}}/E_{\text{sr}}$} &  \multicolumn{1}{c}{$E_{\text{a}}/E_{\text{sr}}$}       &$N$ &   \multicolumn{1}{c}{$E_{\text{droplet}}/E_{\text{sr}}$}      \\
    \hline
    3 &             & -0.1191                & 7 &   -6.544(11)      \\
    4 &  -0.700(4)  & -0.70                  & 8 &   -10.075(16)     \\
    5 & -1.9127(5)  & -1.92                  & 9 &   -14.48(2)       \\
    6 & -3.839(6)   & -3.84                  & 10&   -19.76(4)       \\
    \hline
    \hline
  \end{tabular}
  \label{energytable}
\end{table}
We find that the energy $E_{\text{droplet}}$
scales with the number of pairs, i.e., as $N(N-1)/2$.
This implies that the transition temperature increases linearly with increasing 
$N$.

Since the $N$-body droplet states are only somewhat larger than $r_0$,
the Gaussian interaction model employed
in our work suffers from finite-range effects
and provides only an approximate
description of the $N$-body Efimov scenario.
Note that the recent work by Gattobigio and 
Kievsky~\cite{kievskyarxiv} suggests
a means to correct for these finite-range effects.
Here, we pursue a different approach.
To see how the transition temperature changes
when the droplet energies scale to leading order linearly with
$N$---which is one of the scalings that has been proposed to hold
in the fully universal Efimov scenario~\cite{vonStecherJPB}~\footnote{Note that
  Refs.~\protect\cite{kievskyarxiv,nicholsonprl} propose that the $N$-boson droplet energy
scales as $N(N-1)$ in the universal regime.}---, we apply our combined model to the 
data of Ref.~\cite{vonStecherJPB}.
In that work, the $N$-boson system was assumed to interact through
a combination of two- and three-body potentials.
The resulting transition temperature $T_{\text{tr}}$
is shown by squares in Fig.~\ref{transition}.
The two cases display different large $N$ behavior:
The transition temperature increases roughly
linearly with $N$ for the Gaussian two-body
model interaction but increases
much slower for the system with two- and three-body interactions.
We note that the finite temperature behavior of the trapped
$N=100$ Bose system
was investigated by Piatecki and Krauth using the PIMC 
approach~\cite{KrauthNatComm}.
In the regime where $|E_{\text{trimer}}|$ is much larger than $E_{\text{ho}}$,
Ref.~\cite{KrauthNatComm} 
finds, in agreement with our work,
a transition from a droplet state to a gas-like state.
Reference~\cite{KrauthNatComm} refers to the phase that is governed by the droplet
state as Efimov liquid phase. We emphasize that our calculations neglect decay to
non-universal states. Such states would need to be accounted for if one wanted
to analyze the stability of the droplet phase.

\begin{figure}
\centering
\includegraphics[angle=0,width=0.40\textwidth]{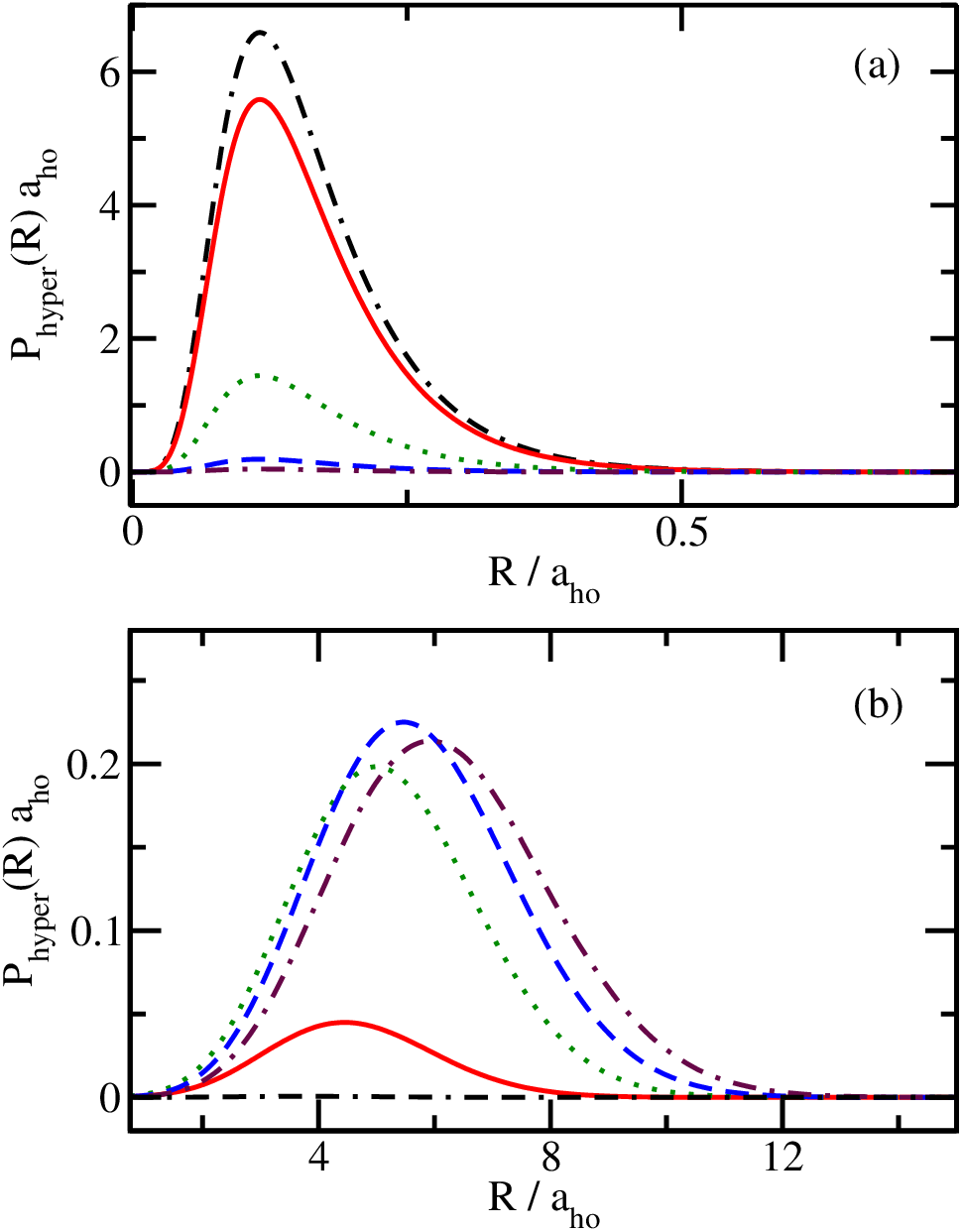}
\caption{(Color online)
  Hyperradial density $P_{\text{hyper}}(R)$ 
for three identical bosons at unitarity
  interacting through $V_{\text{G}}$ with 
$r_0=0.06a_{\text{ho}}$ for various temperatures $T$. 
  Dash-dash-dotted, solid, dotted, dashed, and dash-dotted lines are for 
  $k_BT/E_{\text{ho}}=3, 4, 5, 6,$ and $7$, respectively.
  Panel (a) shows the small $R$ region while panel (b) shows the large $R$
  region.
  Note that panels (a) and (b) have different scales for the $x$-axis and
  the $y$-axis.
 }\label{hyperradialvst}
\end{figure}

We now discuss the system characteristics below
and above $T_{\text{tr}}$ in more detail.
As already mentioned in Sec.~\ref{SecEfimov},
 the hyperradial distribution 
functions $P_{\text{hyper}}(R)$
for the three-boson system interacting through $V_{\text{G}}$
with
ranges $r_0=0.06 a_{\text{ho}}$ and $0.1a_{\text{ho}}$ at low
temperature 
(see Fig.~\ref{hyperradial} for $k_{B}T=0.4 E_{\text{ho}}$)
are essentially identical to the free-space 
three-boson systems with the same $r_0$ at zero temperature.
Figure~\ref{hyperradialvst}
shows the temperature dependence
of $P_{\text{hyper}}(R)$ for $N=3$ and $r_0=0.06a_{\text{ho}}$.
The dash-dash-dotted line shows the hyperradial 
distribution function for $k_BT=3E_{\text{ho}}$, i.e.,
for a temperature below $T_{\text{tr}}$.
For this temperature, $P_{\text{hyper}}(R)$
exhibits a maximum at $R \approx 0.15a_{\text{ho}}$ and falls off
monotonically at larger $R$.
For slightly larger $T$, i.e., $k_B T = 4E_{\text{ho}}$ (solid line), 
the maximum at $R \approx 0.15 a_{\text{ho}}$ is smaller and a second 
peak at $R \approx 4-5 a_{\text{ho}}$ appears.
At yet higher $T$
(above the transition temperature), the amplitude of the large $R$
peak is more pronounced and
the hyperradial distribution function resembles that of 
a gaseous system.
The temperature dependence of the hyperradial distribution function
for the $N=3$ system supports our interpretation introduced above,
namely, the notion that the system undergoes a transition from
an Efimov trimer to a gas state as the temperature changes from
below to above $T_{\text{tr}}$.
The hyperradial distribution functions for larger systems
show analogous behavior, i.e., they support the notion
that the system undergoes a transition from
an $N$-body droplet state to a gas state
with increasing temperature.

\begin{figure}
\centering
\includegraphics[angle=0,width=0.4\textwidth]{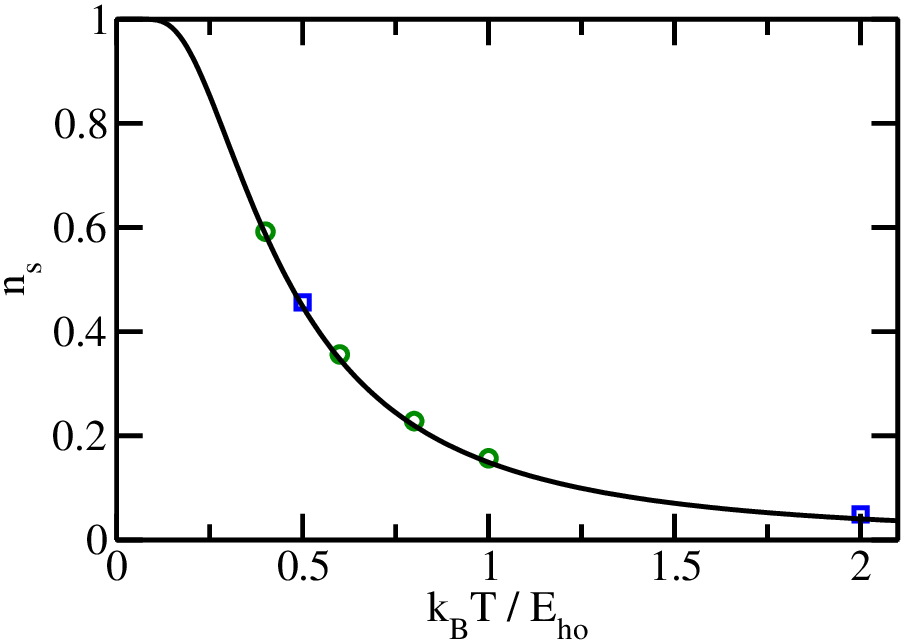}
\caption{(Color online)
  Superfluid fraction $n_s$ as a function of the temperature $T$ 
  for $N$ identical bosons at unitarity.
  The circles and squares show the PIMC results for the Gaussian potential
  $V_{\text{G}}$ 
  with $r_0=0.1a_{\text{ho}}$ for $N=3$ and 
  $4$, respectively.
  The errorbars are smaller than the symbol size.
  For comparison, the solid line shows the result
  obtained using a single-particle model (see text for discussion).
 }\label{3bsf}
\end{figure}

To further characterize the properties of the $N$-boson system,
symbols in Fig.~\ref{3bsf}
show the superfluid fraction $n_s$
as a function of the temperature for $N=3$ and $4$
obtained using the PIMC approach (here,
$r_0=0.1a_{\text{ho}}$ and $1/a_s=0$).
The superfluid fractions for these two system sizes
seem to fall on one curve. 
The solid line, which is obtained analytically (see below
for the model that produces the solid line), provides a good description of
the numerical data. Figure~\ref{3bsf} 
suggests that the superfluid fraction approaches one in the
zero temperature limit and is smaller than $0.05$ for
$ k_B T \gtrsim 2 E_{\text{ho}}$.  
From Figs.~\ref{3benergy} and \ref{345energy} and the surrounding discussion,
we know that the temperature regime  $ k_B T \lesssim 2 E_{\text{ho}}$
is---for the parameters considered---well described by the partition function
$Z_{\text{droplet}}$, i.e., the system behavior is dominated by the
lowest $N$-droplet energy eigen state and its
center of mass excitations.
In particular, this means that the droplet itself can be considered
as ``frozen''.
Correspondingly, we expect that the behavior of the superfluid fraction 
displayed in Fig.~\ref{3bsf} is approximately described by that of a single
harmonically trapped particle of mass $N m_{\text{a}}$
(see the solid line in Fig.~\ref{3bsf}).
We observe that the PIMC points lie 
slightly above the solid line. 
This could be due to the fact that the classical moment of inertia
calculated using the single-particle framework is slightly smaller
than the classical moment of inertia calculated 
using the full Hamiltonian.

We now relate the fall off of the superfluid fraction to the
transition temperature. As discussed above,
the fall off of $n_s$ is governed by center of mass excitations, i.e.,
the relevant temperature scale is set by the harmonic oscillator
frequency. To make some estimates, we say that the superfluid fraction,
defined through the moment of inertia,
is ``undetectably small'' for 
$k_B T$ around $2 E_{\text{ho}}$, independent of the number of particles
and interaction model.
This estimate assumes that the 
absolute value of the eigen energy of the lowest droplet state
is large enough for $Z_{\text{droplet}}$ to provide a
reasonably accurate description of the low temperature dynamics.
For the three- and four-body systems, this implies that
$|E_{\text{droplet}}|$ has to be larger than a few times $E_{\text{ho}}$.
For cold atom systems, the three-body parameter is
found to be approximately 
universal~\cite{PRLbyWang,arXivPaperbyUeda,PRLbyUeda},
i.e., $a_-\approx -9.7 R_{\text{vdW}}$, where $R_{\text{vdW}}$ denotes the van
der Waals length and $a_-$ the scattering length at which the
Efimov trimer merges with the three-atom continuum. 
Using this approximate universality together with the
known relation between $a_-$ and $\kappa_*$~\cite{BraatenHammerReview},
we estimate that 
$E_{\text{trimer}}$
is roughly equal to $-0.024 E_{\text{vdW}}$ at unitarity. Here,
$E_{\text{vdW}}$ is defined as $E_{\text{vdW}}=\hbar^2/(m_{\text{a}}
R_{\text{vdW}}^2)$.
For Cs in a spherically symmetric 
harmonic trap with a frequency  $\nu \approx 2$kHz (a value that can be reached
easily),
the Efimov trimer would have an energy of about $-33 E_{\text{ho}}$ (the
system is approximately described by the circles in Fig.~\ref{3benergy}).
For these
experimental conditions, 
the superfluid fraction is vanishingly small
for $T \gtrsim T_{\text{tr}}$.

A key ingredient of the above analysis is that the fall off of the
superfluid fraction is due to the center of mass excitations.
This suggests an alternative viewpoint that 
defines the superfluid fraction with respect
to the relative degrees of freedom only.
If we replace the $z$-component 
$\hat{L}_{\text{tot},z}$ of the total orbital angular momentum
operator in Eq.~(\ref{eq_quantummomentofinertia}) by the $z$-component
of the relative orbital angular momentum operator
and modify the definition of the
classical moment of inertia accordingly, then
we find that the fall off of the superfluid fraction is correlated with
the transition temperature.
The spirit of the latter approach underlies
the arguments of Ref.~\cite{KrauthNatComm}, which
considers a Bose gas with $N=100$ and refers to the phase governed by the
$N$-droplet state as superfluid phase.
We emphasize, however, that Ref.~\cite{KrauthNatComm} did not
perform any quantitative calculations of the superfluid fraction
or superfluid properties of the system. Instead, Ref.~\cite{KrauthNatComm}
put forward qualitative arguments based on the exchange
paths.

We reiterate that the combined model breaks down when $|E_{\text{droplet}}|$
is not much larger than $E_{\text{ho}}$,
i.e., when the size of the trimer approaches the harmonic oscillator
length.
In this case, the lowest Efimov trimer does not
define a separate energy scale and the phase
transition like feature discussed in this work
disappears.
Qualitatively, we expect that the Bose gas with 
$N=3,4,\cdots$ changes from having a significant superfluid
fraction to a small superfluid fraction as the temperature
increases from zero to a few times $E_{\text{ho}}$.
The $N=100$ case has been considered
in Ref.~\cite{KrauthNatComm}.

\subsection{Single-component gas with a single impurity}
\label{sec_mixture}
This section considers a single-component gas 
consisting of $N-1$ particles with an impurity.
We assume that the impurity interacts with the $N-1$
``background'' atoms through 
the Gaussian potential $V_{\text{G}}$ with diverging
$s$-wave scattering length $a_s$. The background 
atoms do not interact with each other. 
Our goal is to investigate the 
temperature-dependence of the system properties
as the statistics of the $N-1$ background atoms changes
from Bose to Boltzmann to Fermi statistics.
As before, we consider equal mass systems.
Efimov trimers do not exist
for two identical fermions
and a third distinguishable particle (in our case,
the impurity)~\cite{EfimovMassRatio,BraatenHammerReview}. 
For two identical bosons and a third particle or two Boltzmann particles
(i.e., two distinguishable particles)
and a third particle, however, Efimov trimers can exist~\cite{Efimov1973157}.
An interesting question is thus how the finite temperature properties
of the $(N-1,1)$ system with $N \ge 3$ depend on the statistics.

From the discussion in the previous subsection it is clear that
the properties of the trimer
at low temperature determine the characteristics of 
larger Bose systems provided
$|E_{\text{trimer}}|$ is much larger than
$E_{\text{ho}}$.
Throughout this section, we consider the situation where
the lowest energy eigen state of the
$(2,1)$ system with Bose statistics has an
energy
comparable to $E_{\text{ho}}$,
i.e., $|E_{\text{trimer}}| \approx E_{\text{ho}}$ 
[note, the lowest energy eigen state of the
$(2,1)$ system with Boltzmann statistics has the same energy].
For the same model interactions, the lowest energy eigen state of
the $(2,1)$ system with 
Fermi statistics also has an energy comparable to $E_{\text{ho}}$; 
the energy for the system with Fermi statistics is, however, larger
than that for the system with Bose statistics.
We will show that the low temperature properties of the $(N-1,1)$ systems
display, as might be expected naively, statistics dependent characteristics
for temperatures around or below $E_{\text{ho}}$.
Concretely, we 
focus on systems with interspecies Gaussian interactions
with $r_0=0.06a_{\text{ho}}$ and $1/a_s=0$.
The relative ground state energy of the harmonically trapped $(2,1)$ system
with Bose statistics is $0.508 E_{\text{ho}}$ [or $141 \hbar^2/(mr_0^2)$].
For comparison, the relative ground state energy of the corresponding
free-space system is $-18.1 \hbar^2/(m r_0^2)$,
indicating that the trap modifies the lowest energy eigen state
of the free-space system with Efimov characteristics.
The relative ground state energy of the harmonically trapped $(2,1)$ system
with Fermi statistics is $2.785 E_{\text{ho}}$.
The corresponding free-space system is not bound~\cite{blumerev12}.

\begin{figure}
\centering
\includegraphics[angle=0,width=0.45\textwidth]{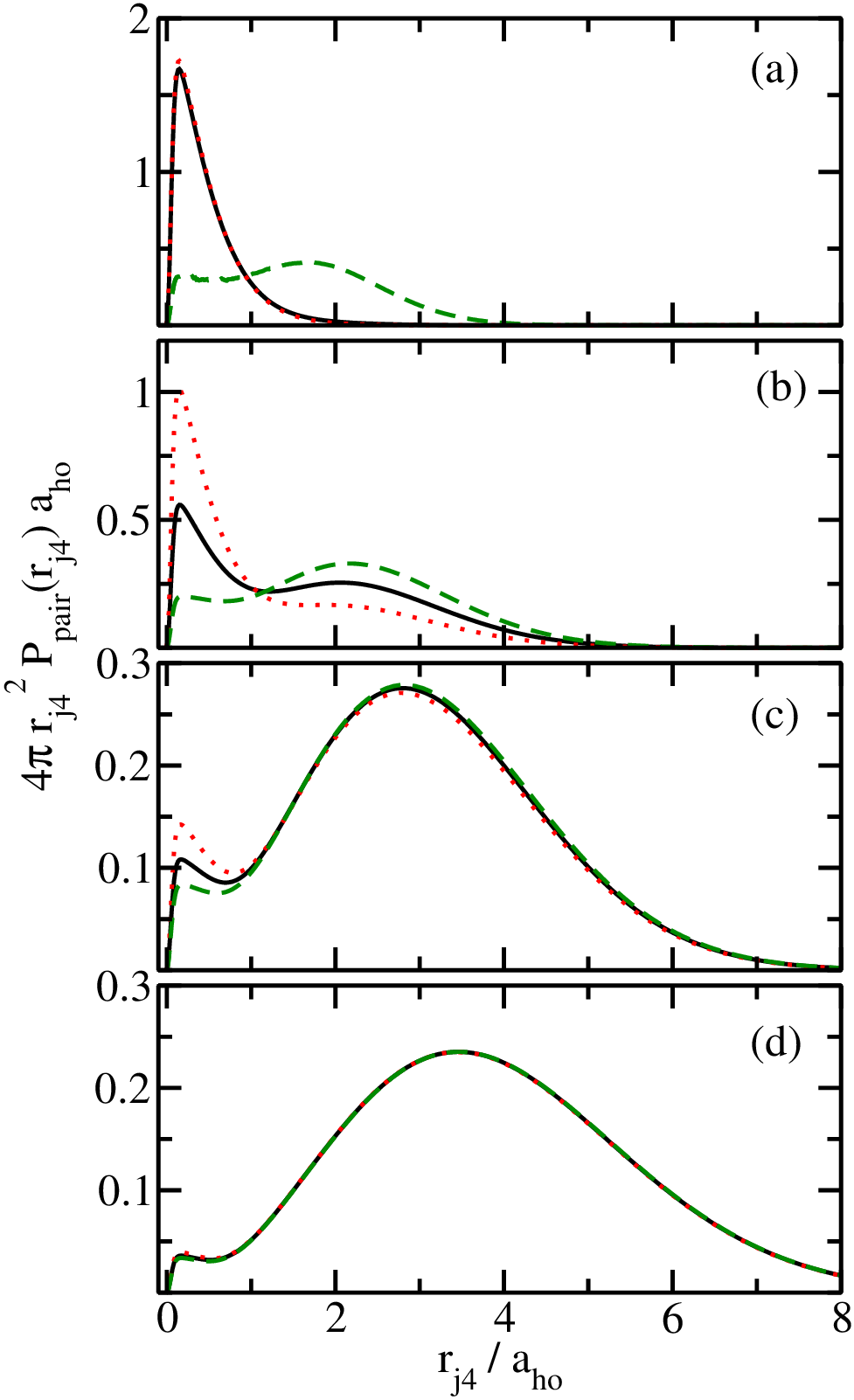}
\caption{(Color online)
  Scaled pair distribution functions $r_{j4}^2 P_{\text{pair}}(r_{j4})$
  ($j<4$)
  for the $(3,1)$ system with interspecies interaction $V_{\text{G}}$
  with $r_0=0.06a_{\text{ho}}$ 
  and diverging interspecies scattering length $a_s$ 
  at temperature (a) $k_B T/E_{\text{ho}}= 0.6$, 
  (b) $k_B T/E_{\text{ho}}= 1.2$, 
  (c) $k_B T/E_{\text{ho}}= 2$, and 
  (d) $k_B T/E_{\text{ho}}= 3$.
  Dashed, solid, and dotted
  lines are for systems with Fermi, Boltzmann, and Bose statistics,
  respectively.
  The error bars are comparable to or smaller than the line widths.
  In panel (a), the solid and dotted lines are hardly distinguishable.
  In panel (d), all three lines nearly coincide.
 }\label{31statistics}
\end{figure}

Figure~\ref{31statistics}
shows the scaled pair distribution functions 
$r_{j4}^2 P_{\text{pair}}(r_{j4})$, $j<4$, 
for the $(3,1)$ system with
$r_0/a_{\text{ho}}=0.06$ and $1/a_s=0$ for different statistics
and temperatures.
The dotted, dashed, and solid lines are for Bose, Fermi, and
Boltzmann statistics, respectively.
Figures~\ref{31statistics}(a)-\ref{31statistics}(d) are for
 $k_BT/E_{\text{ho}}=0.6, 1.2, 2$ and
$3$, respectively.
At high temperature [see Fig.~\ref{31statistics}(d)], the pair distribution
functions are to a very good approximation independent of
the particle statistics.
As the temperature decreases [see Fig.~\ref{31statistics}(c)],
the particle statistics has a visible effect on the pair distribution 
functions.
In the PIMC language,
the temperature in Fig.~\ref{31statistics}(c)
is such that the ``permuted paths'' contribute
only a small fraction 
to the partition function. This implies that the particle statistics
can be treated perturbatively, i.e.,
the partition functions 
$Z_{\text{Bose}}(\beta)$ and $Z_{\text{Fermi}}(\beta)$
of the systems with Bose and Fermi statistics can be
written approximately as
$[Z_{\text{Boltz}}(\beta) \pm \Delta Z(\beta)]/3!$, where
$Z_{\text{Boltz}}(\beta)$ denotes the partition function of the
system with Boltzmann statistics and $\Delta Z(\beta)$ a small correction.
The factor of $1/3!$ arises due to the presence of the three identical
particles (bosons or fermions).
Correspondingly, the sum of the
energies of the
systems with Bose and Fermi statistics equal, to a good
approximation, twice the energy of the system with Boltzmann
statistics.
Indeed, for the temperature considered in
Fig.~\ref{31statistics}(c), we find $E/E_{\text{ho}}=
23.86(2)$, $23.33(2)$, and $22.76(1)$
for Fermi, Boltzmann  and Bose statistics, respectively.
The energy differences are $0.53(4)$ and $0.57(3)$, in
agreement with the expectation based on the perturbative argument.

For yet lower temperatures, the particle statistics becomes
non-perturbative.
In Fig.~\ref{31statistics}(b), e.g., the pair distribution functions
for the three different statistics differ notably.
In Fig.~\ref{31statistics}(a), 
the pair distribution functions for 
the systems with Boltzmann and Bose statistics 
are nearly indistinguishable and notably different from the
pair distribution function for the system with Fermi statistics.
This can be explained as follows. The systems with
Bose and
Boltzmann statistics have the same ground state energy
while the system with Fermi statistics has a notably larger ground state energy.
Due to the absence of bound trimer states 
for the system with Fermi statistics for vanishing confinement (i.e., for $w=0$) , the pair distribution
function
is fully determined by the trap length and the temperature~\cite{stringariFermireview,blochreview,blumerev12}.
For the systems with Bose and Boltzmann statistics, the
pair distribution function takes on large values at small $r$,
reflecting the fact that these systems form a droplet like state for vanishing
confinement.
An important consequence is that the two-body contacts
for the systems with Bose and Boltzmann statistics are,
in the low temperature regime,
much larger than the two-body contact
for the system with Fermi statistics.

\begin{figure}
\centering
\includegraphics[angle=0,width=0.45\textwidth]{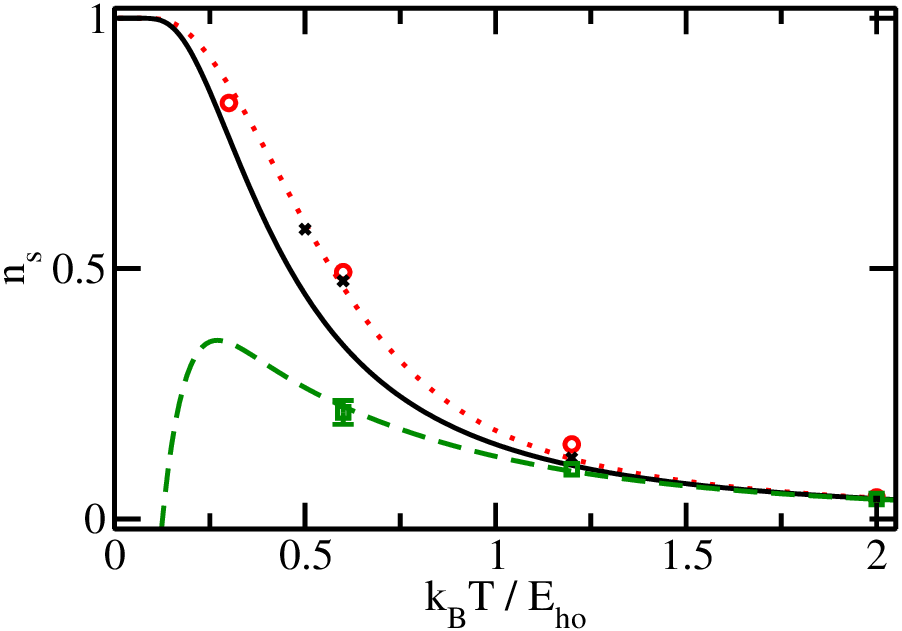}
\caption{(Color online)
  Superfluid fraction $n_s$ as a function of the temperature $T$ for the $(3,1)$ system
  with interspecies potential
  $V_{\text{G}}$ with $r_0=0.06a_{\text{ho}}$ and $1/a_s=0$. 
  The circles, crosses, and squares are obtained from the PIMC simulations with Bose,
  Boltzmannn, and Fermi statistics, respectively.
  The error bars are only shown when they are larger than the symbol size.
  For comparison, dotted, solid, and dashed
  lines show the superfluid fraction for the non-interacting $(3,1)$ systems with Bose, Boltzmann, and Fermi statistics,
  respectively.
 }\label{31sf}
\end{figure}

Symbols in Fig.~\ref{31sf} 
show the superfluid fraction $n_s$ as a function of the
temperature for the $(3,1)$ system with interspecies potential $V_{\text{G}}$ with $r_0=0.06a_{\text{ho}}$ and infinitely
large $s$-wave scattering length. Circles, crosses, and squares are for Bose,
Boltzmann, and Fermi statistics, respectively.
As the temperature decreases, the superfluid fraction increases for 
 the systems with Boltzmann and Bose statistics and reaches 
$1$ at zero temperature.
The superfluid fraction of the $(3,1)$ system with Fermi statistics lies below
that for the $(3,1)$ system with Bose and Boltzmann statics at high
temperature. Our calculations go down to $k_BT=0.6E_{\text{ho}}$. Based on our
earlier work~\cite{OurArxivPaper}, we expect that the superfluid fraction for
the system with Fermi statistics will
take on negative values as the temperature approaches zero.
At high temperature, the perturbative analysis, introduced earlier for the
energy, can be applied to the 
superfluid fraction. The ``permuted paths'' contribute perturbatively 
to the quantum moment of inertia and the classical moment of inertia. The combination of the two
gives rise to a correction of the superfluid fraction calculated from the
``unpermuted paths'', i.e., a correction to the superfluid fraction for the
$(3,1)$ systems with Boltzmann statistics due to the exchanges of identical
particles.
At $k_BT=2E_{\text{ho}}$, we find $n_s= 0.03976(5),$ $0.04132(1)$,
and $0.04294(3)$ for the $(3,1)$ systems with Fermi, Boltzmann, and Bose statistics, respectively.
The differences are $0.00156(6)$ and $0.00162(4)$, in agreement with the
expectation based on the perturbative argument. 
For comparison, dotted, solid, and dashed lines show the superfluid fraction
for the non-interacting (3,1) systems with  Bose, Boltzmann 
and Fermi statistics, respectively.
For the system with Bose statistics, 
the unitary interactions change the superfluid fraction only slightly.
For the system with Boltzmann statistics, the interactions have a notably
larger effect on the superfluid fraction.
The non trivial shift comes from the interplay between the temperature and
the interactions.

Finally, we comment that the single-particle model, where the droplet is
describled as a single particle of mass $N m_{\text{a}}$, is not applicable.
The superfluid fraction for this model coincides with the solid line in
Fig.~\ref{31sf}. If $|E_{\text{trimer}}|$ was much larger than
$E_{\text{ho}}$, we would expect that the superfluid fraction for the systems with
Bose and Boltzmann statistics would follow the solid line. The fact that the
symbols deviate from the solid line indicates that the single particle model
is not applicable. Interestingly though, the superfluid fracction seems to
only change weakly as $E_{\text{trimer}}/E_{\text{ho}}$ changes, suggesting
that $n_s$ is not a sensitive probe of the phase transition like feature or
absence thereof.

\section{Conclusions}
\label{sec_conclusion}
This paper considered 
the finite temperature properties of small $s$-wave interacting 
systems under spherically symmetric
harmonic confinement. 
For two particles in the harmonic trap, we compared 
the condensate 
and superfluid fractions as a function of the temperature.
The role of the particle statistics on these quantities
was discussed.
For two Boltzmann particles, the
condensate fraction exhibits a strong dependence on the interaction
strength while
the  superfluid fraction
is only weakly dependent on the interaction strength.
Changing from Boltzmann to Bose statistics changes the observables
by a relatively small amount while changing from Boltzmann to
Fermi statistics introduces significant quantitative changes.

We further considered $N$-bosons with
finite-range two-body Gaussian interactions at unitarity in the regime where
the absolute value of the $N$-boson droplet energy $|E_{\text{droplet}}|$ is
much larger than the harmonic oscillator energy.
We observed a sharp transition as the temperature
increases from a liquid droplet like state to 
a gas-like state. The energy, heat capacity, 
hyperradial distribution
function, and 
superfluid
fraction were monitored as
a function of the
temperature. A simple model that semi-quantitatively
captures the entire temperature regime
was proposed.
The model was not only applied to systems with Gaussian interactions but
also to systems with two- and three-body
interactions.
No evidence for ``intermediate phases'' such as a gas consisting of trimers or
tetramers was found.
Finally, we considered the $(3,1)$ system with infinitely
large interspecies scattering length.
We compared the pair distribution
function for systems with Bose, Boltzmann, and Fermi statistics.
We established that the statistics can be treated perturbatively
at high temperature.

In the future, it will be interesting to extend the few-body studies presented
here to larger number of particles. For bosons, this should be fairly
straightforward. For fermions, however, the sign problem will place
constraints on the temperature regime that can be covered. An important
question is if the $N$-boson droplet state discussed here can be probed
experimentally. Our calculations excluded non-universal energetically lower
lying states, which could lead to atom losses. Moreover, we assumed that the
system is in thermal equilibrium. In practice, experimental investigations
will have to work in a parameter regime where the equilibration time is faster
than the atom loss time. It remains an open question if quench experiments
such as those recently conducted at JILA~\cite{makotyn2014universal} could, if applied to small systems,
probe the phase transition like feature discussed in this work.
A possible scheme would be to start with a weakly interacting system with
known but variable temperature, to jump the magnetic field to unitarity, and
last to probe the system after a variable hold time.

Our calculations for few-fermion systems showed that a temperature of less
than $E_{\text{ho}}/k_B$ leads to notable changes in the structural properties.
This suggests that the analysis of few-fermion experiments has to account for
finite-temperature effects. A similar conclusion was reached in
Refs.~\cite{lewensteinpra13,2014arXiv1406.0400S}, which considered---motivated by the Heidelberg
experiments~\cite{jochimscience11,jochimprl12,jochimscience13}---the temperature dependence of one-dimensional
few-fermion systems.

{\em{Acknowledgement:}}
Support by the National
Science Foundation (NSF) through Grant No.
PHY-1205443
is gratefully acknowledged.
This work used the Extreme Science and Engineering
Discovery Environment (XSEDE), which is supported by
NSF Grant No. OCI-1053575, and the
WSU HPC.
YY and DB acknowledge support from the Institute for Nuclear Theory
during the program  INT-14-1,
``Universality in Few-Body Systems: Theoretical
Challenges and New Directions''.

%\bibliographystyle{prsty}
%\bibliography{mybib}
%merlin.mbs apsrev4-1.bst 2010-07-25 4.21a (PWD, AO, DPC) hacked
%Control: key (0)
%Control: author (8) initials jnrlst
%Control: editor formatted (1) identically to author
%Control: production of article title (-1) disabled
%Control: page (0) single
%Control: year (1) truncated
%Control: production of eprint (0) enabled
%
\end{document}